\documentclass[amssymb,prb,showpacs,twocolumn]{revtex4-1}

\usepackage{color,graphicx,hyperref,mathptmx,mathtools}
\usepackage[applemac]{inputenc}

\graphicspath{{figures/}}

\def\vec#1{{\bf #1}}

\begin{document}

\title{N\'eel temperature and reentrant H-T phase diagram of quasi-2D frustrated magnets}

\author {Burkhard Schmidt and  Peter Thalmeier}

\affiliation{Max Planck Institute for the Chemical Physics of Solids, 01187 Dresden, Germany}

\date{\today}

\begin{abstract} In quasi-2D quantum magnets the ratio of N\'eel temperature $T_\text N$ to Curie-Weiss temperature $\Theta_\text{CW}$ is frequently  used as an empirical criterion to judge the strength of frustration. In this work we investigate how these quantities are related in the canonical quasi-2D frustrated square or triangular $J_1$-$J_2$ model. Using the self-consistent Tyablikov approach for calculating $T_\text N$ we show their dependence on the frustration control parameter $J_2/J_1$ in the whole N\'eel and columnar antiferromagnetic phase region. We also discuss approximate analytical results. In addition the field dependence of $T_\text N(H)$ and the associated possible reentrance behavior of the ordered moment due to quantum fluctuations is investigated. These results are directly applicable to a class of quasi-2D oxovanadate antiferromagnets. We give clear criteria to judge under which conditions the empirical frustration ratio $f=\Theta_\text{CW}/T_\text N$ may be used as measure of frustration strength in the quasi-2D quantum magnets.
\end{abstract} 


\maketitle

\section{Introduction}
\label{sect:introduction}

Long range magnetic order is prevented at finite temperature in strictly 2D spin systems with a continuous symmetry~\cite{mattis:06}. Commonly the susceptibility would not have a singular cusp at a finite temperature but show a broad maximum at a temperature that corresponds roughly to the average energy scale $J_\text c$ of the intra-plane exchange interactions. This behavior is indeed found experimentally  in quasi-2D magnets and is also obtained theoretically using e.\,g. finite temperature Lanczos method (FTLM) based on exact diagonalization of finite clusters~\cite{shannon:04,schmidt:15}. However in reality these magnets nevertheless mostly exhibit long range magnetic order at even lower temperature. This is due to their quasi-2D character caused by the finite inter-plane interactions $J_\perp\ll J_\text c$ in real compounds such as the $S=1/2$ layered vanadium compounds~\cite{melzi:00,carretta:04,kaul:04,kini:06} listed in Table \ref{tbl:exchange}. A famous example is La$_2$CuO$_4$, the antiferromagnetic parent compound of high-$T_\text c$ superconductors. Although the inter-plane coupling is extremely small $J_\perp/J_\text c\approx 1.3\cdot10^{-6}$ a large Néel temperature $T_{\text N} = 325\; \text{K}$  is observed~\cite{thio:88}. This is due to the fact that in quasi-2D magnets the ordering temperature is still determined by the large intra-plane exchange ($J_\text c\approx 116\;\text{meV}$) and is only logarithmically suppressed roughly by the factor $\ln(J_\text c/J_\perp)$. The physical reason is that a strictly 2D Heisenberg system is at a quantum critical point with algebraic decay of long range correlations. Then even tiny  interlayer coupling may lead to sizable 3D ordering temperature~\cite{majlis:92}.

This matter is well understood in the nearest-neighbor (n.n.) Heisenberg antiferromagnet and has been quantitatively investigated with numerical MC simulations~\cite{yasuda:05} and approximate theories based on Tyablikov RPA theory~\cite{tyablikov:67, yablonskiy:91,majlis:92,majlis:93} and also more advanced analytical methods~\cite{irkhin:97,ihle:99,katanin:12,furuya:16}. On the other hand the restriction to only n.n.{} interactions which are furthermore isotropic in the lattice misses a large body of known frustrated quasi-2D magnets that are described by the square lattice  $J_1$-$J_2$ model or the related anisotropic triangular  $J_1$-$J_2$ models (Fig.~\ref{fig:lattices}). In these systems the general behavior of the ordering temperature $T_\text N(\phi,J_\perp)$ as function of frustration control parameter $\phi=\tan^{-1}J_2/J_1$ has not been investigated systematically in the two possible N\'eel (NAF) and columnar (CAF) antiferromagnetic regions (inset of Fig.~\ref{fig:frustration}) but in the frustrated FM case~\cite{hartel:10}. In the interior of the AF phase regions it is well understood how the ordered moment reduction at zero temperature depends on $\phi$, e.g. from linear spin wave theory (LSW) and comparison with exact diagonalization (ED) results~\cite{schmidt:11,schmidt:17}. The ordered moment is determined by the interplay of quantum fluctuations and frustration and may be completely suppressed  on approaching small intervals of $\phi$ or $J_2/J_1$ around the classical phase boundaries where a spin liquid state or more exotic order is expected and LSW breaks down. The frustration dependence of the ordered moment will lead to a concomitant dependence of the overall energy scale of spin excitations. Consequently the quasi-2D finite N\'eel temperature should show similar strong dependence on the degree of frustration. This is often empirically characterized by a `frustration ratio' $f=\Theta_\text{CW}/T_\text N$ where $\Theta_\text{CW}$ is the Curie-Weiss temperature. This ratio is expected to become large in the strongly frustrated regime where magnetic order breaks down and $T_\text N$ vanishes. This may, however, not be the only possible origin for a large $f$ value. On the other hand it is also useful to define a microscopic frustration ratio $\kappa(\phi)$ which characterizes how far the ground state energy of fundamental frustrated square and triangular tiles is increased with respect to  their unfrustrated constituents.

It is the purpose of this work to clarify the connection between the quantities characterizing the frustrated magnet ground state and its finite temperature behavior.  In particular  we discuss how the size of the interlayer coupling $J_\perp$ can be estimated from the experimentally determined values of $T_\text N$ and $J_1$, $J_2$. This is of great practical importance for frustrated magnets and we show how this may be achieved for the well-investigated oxovanadate layered compounds. For this purpose we use the simple analytical Tjablikov theory which is based on a self-consistently scaled spin wave dispersion. We extend this approach to calculate the field dependence of $T_\text N$ which may be nonmonotonic due to the field-induced suppression of quantum fluctuations. Accordingly a reentrant behavior for the ordered moment and a reentrant $H$-$T$ phase diagram may be derived and we discuss a realistic example. 

\section{Square and anisotropic triangular frustrated exchange model and their classical and quantum phases}
\label{sect:model}

\begin{figure}
	\includegraphics[width=.55\columnwidth]{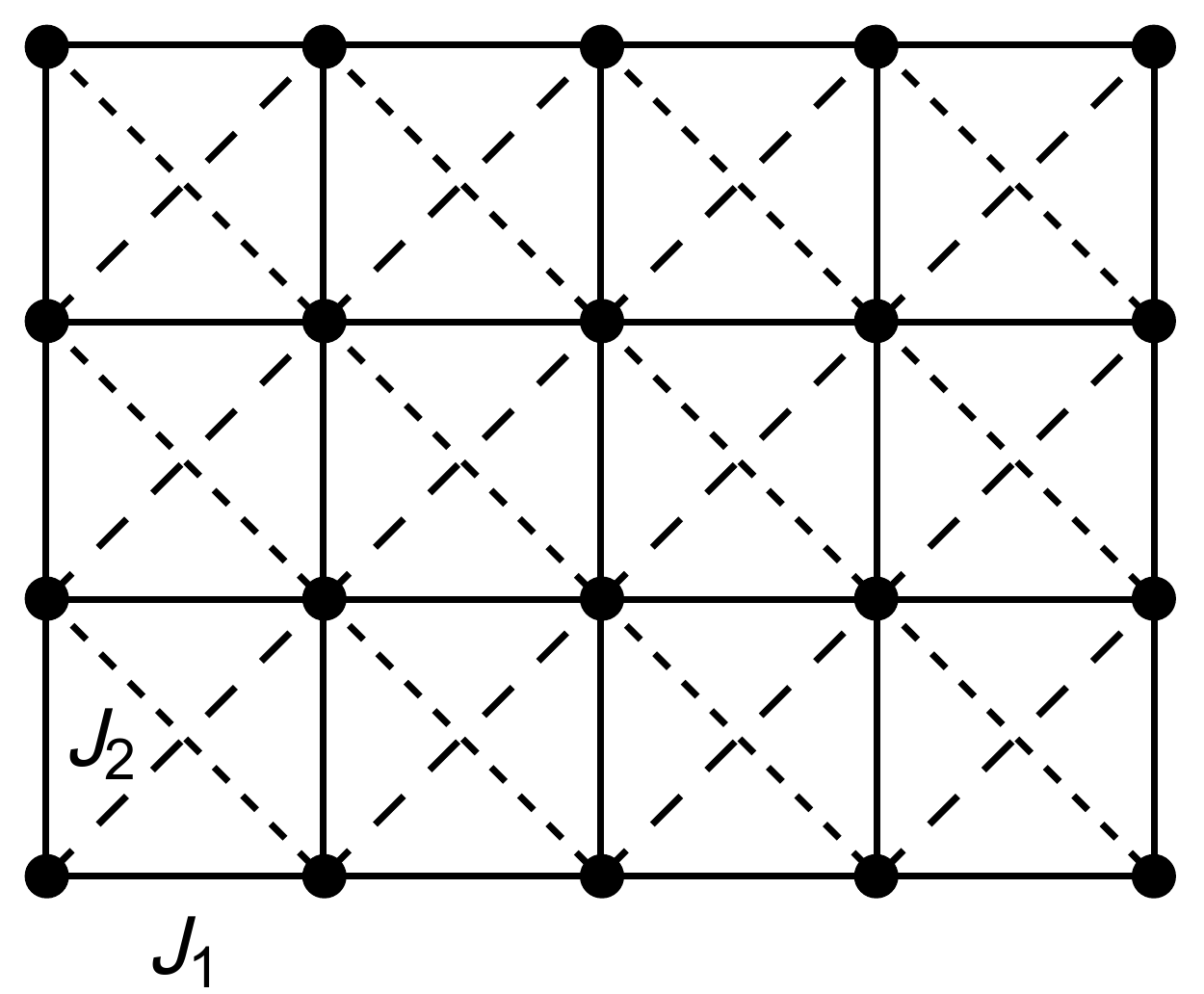}\\
	\includegraphics[width=.65\columnwidth]{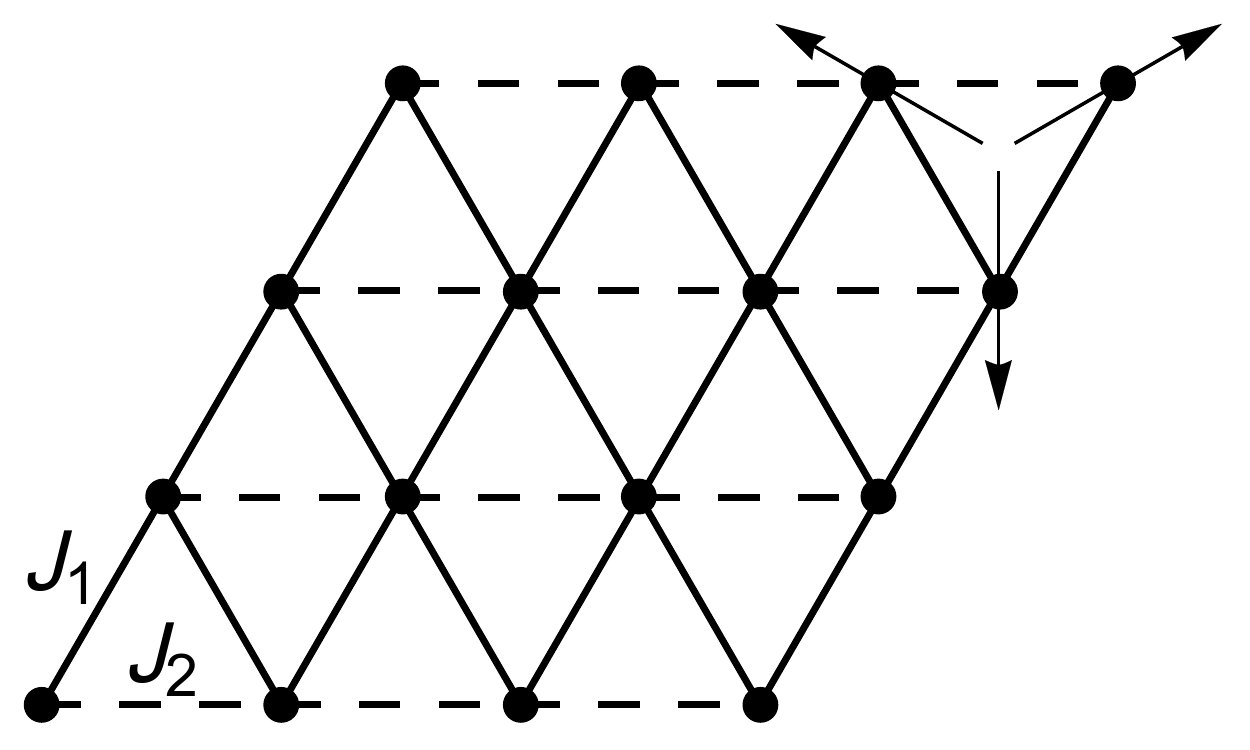}
	\caption{$J_1$-$J_2$ exchange models on the square (a) or anisotropic triangular (b) lattices. In (b) the spin configuration of the maximally frustrated $120^\circ$ structure $(\triangle)$ is indicated. In (a) $J_1$ and $J_2$ denote  isotropic n.n. and n.n.n. interactions, in (b) they denote bond-anisotropic n.n. interactions}
	\label{fig:lattices}
\end{figure}
These models provide a most instructive insight into the essentials of frustrated magnetism~\cite{schmidt:17}. Furthermore numerous realizations in magnetic compounds exist that allow for a comparison of theoretical to experimental results. Here we employ the generic Heisenberg $J_1$-$J_2$ exchange model  for both lattices as illustrated in Fig.~\ref{fig:lattices}~\footnote{In this and all other multi-part figures we use the convention that panels are labeled a, b, c, \ldots from left to right and/or from top to bottom.}:
\begin{equation}
    {\cal H}_\text{ex}
    =
   J_1 \sum_{\left\langle ij\right\rangle_1}
    \vec S_i\cdot\vec S_j+
    J_2 \sum_{\left\langle ij\right\rangle_2}
     \vec S_i\cdot\vec S_j
    \label{eqn:hex12}
\end{equation}
It has the attractive property of having just one control parameter, the frustration ratio $J_2/J_1$ which allows to tune through a rich phase diagram in both cases. It is convenient to use a polar parametrization of the model which maps to a control parameter $\phi\in [-\pi,\pi]$ in a compact interval according to 
\begin{eqnarray}
    J_{1}=J_{\text c}\cos\phi,
    &\quad&
    J_{2}=J_{\text c}\sin\phi,
    \nonumber \\
    J_{\text c}=\sqrt{J_{1}^{2}+J_{2}^{2}},
    &\quad&
    \phi=\tan^{-1}\left(\frac{J_{2}}{J_{1}}\right)
     \label{eqn:exchange}
\end{eqnarray}
We note that the anisotropic triangular model of Fig.~\ref{fig:lattices}(b) can be obtained from (a) by tilting the lattice and cutting one of the diagonal $J_2$ exchange bonds. Therefore, while (a) is an interaction frustrated model with n.n. and n.n.n bonds (b) is a geometrically frustrated model with only (real space anisotropic) n.n. bonds. The classical phase diagram is obtained from the minimum of the classical ground state energy $E_\text{cl}=NS^2J_{\vec Q}$ where $\vec Q$ is the magnetic ordering vector and the exchange function is given by
\begin{equation}
    J_{\vec k}=
    \left\{
      \begin{array}{ll}
	   \boxtimes: & J_{1}(\cos k_x+\cos k_y)+2J_2\cos k_x\cos k_y+J_\perp\cos k_z \\
	   \triangle: & 2J_{1}\cos\frac12k_x\cos\frac{\sqrt{3}}{2}k_y +J_{2}\cos k_x+J_\perp\cos k_z 
      \end{array}
    \right.
 \label{eqn:exfourier}   
\end{equation}
for square ($\boxtimes$) and triangular ($\triangle$) lattices, respectively. The symbols for the special cases of the $J_1$-$J_2$ exchange model are defined in Table~\ref{tbl:symbols}.

Here we included already the small AF coupling $J_\perp >0$ between the 2D layers which are placed on top of each other to mimic the quasi-2D magnetism of real compounds. The moments are then staggered perpendicular to the 2D planes such that $Q_z=\pi$ for the 3D ordering vector $\vec Q=(Q_x,Q_y,Q_z)$. Three classical in-plane 2D phases ($J_\perp=0$) occur in the same regions of $\phi$ for square and triangular lattice: Ferromagnetic (FM) for $\phi\in [0.85\pi,-0.5\pi]$ with $(Q_x,Q_y)=(0,0)$,  N\'eel antiferromagnet (NAF) for $\phi\in [-0.5\pi,0.15\pi]$  with  $(Q_x,Q_y)=(\pi,\pi)$
and for  $\phi\in [0.15\pi,0.85\pi]$ either a columnar antiferromagnet (CAF) for square lattice with  $(Q_x,Q_y)=(\pi,0)$ , $(Q_x,Q_y)=(0,\pi)$ or a spiral phase (SPI) for triangular lattice with $(Q_x,Q_y)$ varying continuously as function of $\phi$ between NAF and FM case~\cite{schmidt:14}.

\begin{figure}
	\includegraphics[width=.9\columnwidth]{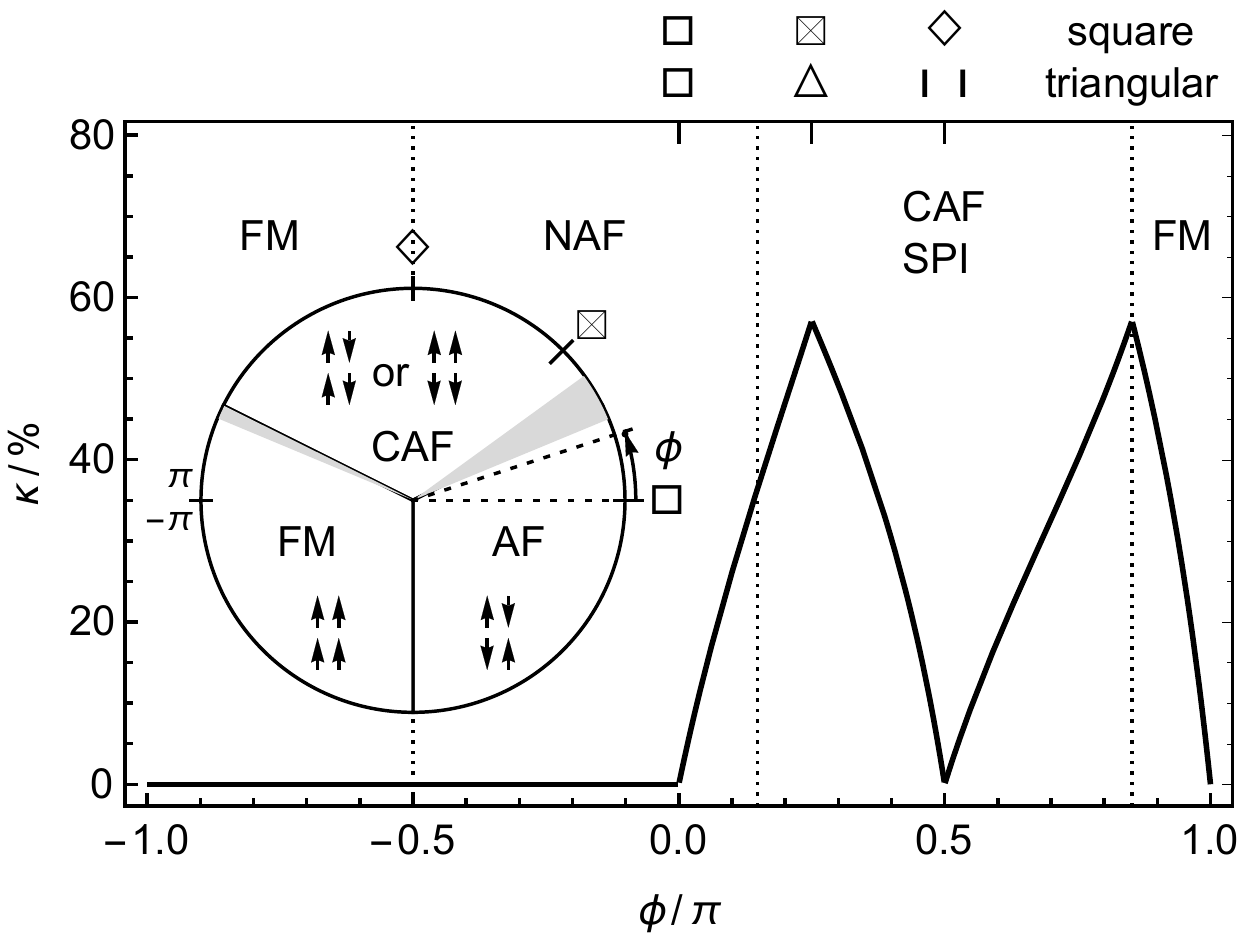}
	\caption{Frustration degree of $J_1$-$J_2$ model in percent (Eq.~(\ref{eqn:neel:kappa})). The maximum appears for $J_1=J_2$ and at CAF(SPI)/NAF phase boundary. Here and in the following figures the symbols on top are defined in Table~\ref{tbl:symbols}. Inset: Classical phase diagram of square and anisotropic triangular $J_1$-$J_2$ model  in polar presentation ($-\pi\leq\phi\leq\pi$, $J_\text c=1$). The CAF phase is twofold degenerate $(a,b)$. For the triangular model it is replaced 
by the generally incommensurate spiral (SPI) phase.}
	\label{fig:frustration}
\end{figure}
The classical ordered moment $m_{\vec Q}=S$ is only realized in the FM phase. In the AF or SPI phases quantum fluctuations strongly reduce the moment, depending on the size of the spin S and frustration control parameter $\phi$~\cite{schmidt:14}. This may be concluded from linear spin wave (LSW)~\cite{shannon:04,schmidt:14} as well as unbiased numerical exact diagonalization (ED) analysis on finite tiles~\cite{schmidt:11}. At the classical phase boundaries NAF/CAF or NAF/SP  ($\phi=0.15\pi)$ and CAF/FM or SPI/FM  ($\phi=0.85\pi)$ the quantum reduction of the moment diverges and long range magnetic order is destroyed. The leads to the possibility of  much discussed `spin liquid' phases reviewed in Refs.~\onlinecite{starykh:15,savary:17,schmidt:17}. This designation is used generically for many-body ground states that do not have long range magnetic order but rather exhibit finite range or algebraic spin correlations or show a more exotic order like valence bond solid or spin nematic state~\cite{schmidt:17}.

In this work our main goal is the analysis of the overall variation of ordering temperature $T_\text N(\phi,J_\perp)$ as function of control parameters in the {\em magnetically\/} ordered phases which dominate the phase diagram, using the linear spin wave (LSW) theory. The range of $\phi$ values where spin wave theory predicts the vanishing of ordered moments and becomes unreliable corresponds approximately to the narrow regions around  $\phi/\pi\approx 0.15, 0.85$ where possibly a dimer spin liquid and a spin nematic phase appear, respectively. Numerous other analytical and numerical methods have been used to investigate this strongly frustrated region (Fig.~\ref{fig:frustration}), e.g. in Refs.~\onlinecite{weihong:99,yunoki:06,weng:06,heidarian:09,hauke:10,schmidt:11,jiang:12}.

The obtained $J_2/J_1$- or $\phi$-intervals of the spin liquid phase depend strongly on the method used (see Table 4 in Ref.~\onlinecite{schmidt:17}), therefore the precise value of upper and lower boundary of the spin liquid interval is an open question. Its absolute width as compared to the magnetic regions ($\Delta\phi/\pi\simeq 0.6$) is, however, quite small, e.g. from exact diagaonlization (ED) with scaling analysis for the square lattice model one obtains $\Delta\phi/\pi\simeq 0.075$ for the spin dimer phase interval and $\Delta\phi/\pi\simeq 0.020$ for the spin nematic interval, indicated by the grey-shading in the inset of Fig.~\ref{fig:frustration}. It is not clear to which extent the above methods for the spin liquid regimes are able to include the effect of finite temperature and interlayer coupling. The latter may indeed further shrink the spin liquid phase interval by stabilizing magnetic order. One should note that various other additional interactions which may destabilize the spin liquid sectors~\cite{schmidt:17} so that to achieve this ground state fine tuning of exchange parameters is necessary.

Given this situation we restrict here to the linear spin wave method because there it is known how a self consistent theory at finite temperature may be obtained empirically to calculate the ordering temperature. However one should be aware that for $\phi$ inside the (not well known) spin liquid intervals the depression of the 3D ordering $T_\text N$ is only qualitatively described by spin wave theory and in reality may even be more rapid when approaching the center of the interval. In any case our interest here is focused on the stable magnetic regions in the phase diagram. And there are indeed plenty of known ordered quasi-2D magnets  described by the $J_1$-$J_2$ model, one extended class will be discussed in Sec.~\ref{sec:oxo}. On the other hand there is so far no compound example that realizes a spin liquid phase of the (anisotropic) triangular or square lattice, therefore our focus on the magnetically ordered regime is empirically justified.

The quantum suppression of the ordered moment $m_{\vec Q}(\phi)$ shows considerable variation with $\phi$ inside the magnetic phase region and a continuous suppression to zero from both sides when approaching the quantum phase transition to the narrow spin liquid sectors. This is found from both LSW and ED~\cite{schmidt:11}, DMRG~\cite{jiang:12}, dimer series expansion~\cite{weihong:99}  and many other techniques reviewed in Ref.~\onlinecite{schmidt:17}. This naturally suggests that the actual ordering temperature $T_\text N(\phi)$ of quasi-2D systems also shows considerable variation with $\phi$ inside the large NAF and CAF phase regions and vanishes continuously when $\phi$ approaches the narrow spin liquid regimes from both sides. However there is no analysis of  $T_\text N(\phi)$ in the whole NAF and CAF phase sectors available. So far mostly the  $J_\perp/J_\text c$ dependence of the {\it unfrustrated} ($\phi=0$) AF has been investigated~\cite{majlis:92,siurakshina:00,yasuda:05}. But $T_\text N(\phi) $ is an important practical issue because firstly, many known frustrated $J_1-J_2$ type compounds belong to these sectors and secondly the experimental value of T$_N$  compared to the paramagnetic Curie-Weiss temperature $\Theta_{\text CW}$ is usually taken as an empirical indicator of the strength of frustration in a magnet~\cite{obradors:88,ramirez:94,wilfong:17}.
\begin{table}
    \centering
	\begin{tabular}[c]{llll}
	symbol & exchange & $\phi/\pi$  & model \\
	\hline
	 $\boxtimes$ & $J_1,J_2$ & any & general frustrated model \\
	 $\square$ & $J_2=0$ & 0 & pure N\'eel, latt. const. $a$ \\
	$\Diamond$ & $J_1=0$ & $\pi/2$ & pure N\'eel, latt. const. $\sqrt{2}a$\\
	\hline
	$\Delta$ & $J_1,J_2$ & any & anisotropic triangular \\
        $\triangle$ & $J_1=J_2$ & $\pi/4$ & isotropic triangular \\
        $\square$ & $J_2=0$ & 0 & pure N\'eel, latt. const. a \\
         $\parallel$ & $J_1=0$  & $\pi/2$ & decoupled 1D AF chains  \\
	\end{tabular}
    \caption
    {Definition of symbols for general and special exchange models for square (top three) 
    and triangular lattice (bottom four). 
     }
    \label{tbl:symbols}
\end{table}

\section{Empirical frustration parameter and microscopic frustration degree}

For magnetic materials, in many cases two parameters are easily accessible experimentally, the paramagnetic Curie Weiss temperature and the N\'eel (or Curie) temperature of the ordered phase. At high temperatures where moments become decoupled the uniform susceptibility is described by the empirical expression
\begin{equation}
\chi=\frac{C}{T+\Theta_\text{CW}}
\end{equation}
where $C$ is a constant and $\Theta_\text{CW}$ is the Curie-Weiss temperature which is positive or  negative for AF or FM  materials, respectively. It  is defined through the first term of the high temperature series expansion (HTSE) of $\chi(T)$~\cite{schmidt:15} according to
\begin{eqnarray}
	\chi
	&=&
	\frac{S(S+1)}3\beta J_\text c(1-\beta k_\text B\Theta_\text{CW}),
	\\
	\Theta_\text{CW}
	&:=&
	\frac{S(S+1)}{3k_\text B}\sum_nJ_{ii+n}
	\nonumber\\
	&=&
	\frac{2S(S+1)}{3k_\text B}J_{\vec k=0}
\end{eqnarray}
where the susceptibility per site $i$ is given in units of $\chi_0=\mu_0(g\mu_\text B)^2/J_\text c$. Explicitly, for the 3D model we have
\begin{equation}
    k_\text B\Theta_\text{CW}=
    \left\{
      \begin{array}{l@{\qquad}l}
	   \boxtimes :\; J_1+J_2+\frac12J_\perp   \\
	   \triangle :\;  J_1+\frac12J_2+\frac12J_\perp
      \end{array}
    \right.
 \label{eqn:TCW}   
\end{equation}
On the basis of a mean field (MF) approximation $\Theta_\text{CW}$ is frequently associated with the AF ordering or N\'eel temperature $T_\text N$ (the second experimental parameter)~\footnote{Indeed for ferromagnets, $|\Theta_\text{CW}|$ is identical to the mean-field Curie temperature $T_\text c$}.  For the 3D (simple tetragonal or hexagonal) model  the MF values are given by
\begin{equation}
	T_\text N=\frac{2S(S+1)}{3k_\text B}|J_{\vec Q}|
\end{equation}
This also means that the mean field N\'eel temperature ($S=1/2$) is equal to the classical ground state energy per bond according to $k_\text BT_\text N=|E_\text{cl}|/(N/2)$. Explicitly one obtains
\begin{equation}
    k_\text B T_{N}
    =
    \left\{
      \begin{array}{l@{\qquad}l}
	   \boxtimes\;\ \text{NAF} :\; J_1-J_2+\frac12J_\perp
	   \vspace{0.2cm}\\
	   \boxtimes\;\ \text{CAF} :\; J_2+\frac12J_\perp
	   \vspace{0.2cm}\\
	   \triangle\;\ \text{NAF} :\; J_1-\frac12J_2+\frac12J_\perp
	   \vspace{0.2cm}\\
	   \triangle\;\ \text{SPI} :\;  \frac12J_2\left[
	       1+\frac12(J_1/J_2)^2
	   \right]+\frac12J_\perp
      \end{array}
    \right.
 \label{eqn:mftn}   
\end{equation}
For the unfrustrated ($J_2=0$)  NAF phases ($\boxtimes, \triangle$) evidently $\Theta_\text{CW}=T_\text N$, for any $J_\perp$ in mean field approximation. Naturally the mean field expressions in Eq.~(\ref{eqn:mftn}) cease to be reasonable for $J_\perp/J_\text c\ll 1$ where $T_\text N$ has to approach zero for the 2D lattice.

For frustrated magnets (geometrically or interaction frustrated)  intuitively the temperature for long range order should be suppressed because of the competition between exchange bonds whose exchange energy cannot be minimized simultaneously for all bonds, i.\,e. one would expect $T_\text N \ll \Theta_\text{CW}$ in strongly frustrated systems. Therefore it has become customary in experimental investigations to characterize frustrated magnets by the ratio
\begin{equation}
f :=\frac{\Theta_\text{CW}}{T_\text N}
\end{equation}
With $T_\text N$ strongly suppressed one would then obtain $|f| \gg 1$ with the sign given by that of $\Theta_\text{CW}$. Thus $f$ might be regarded as a direct measure of the degree of frustration in a particular magnet~\cite{obradors:88, ramirez:94,wilfong:17}.

\begin{figure}
	\includegraphics[width=.9\columnwidth]{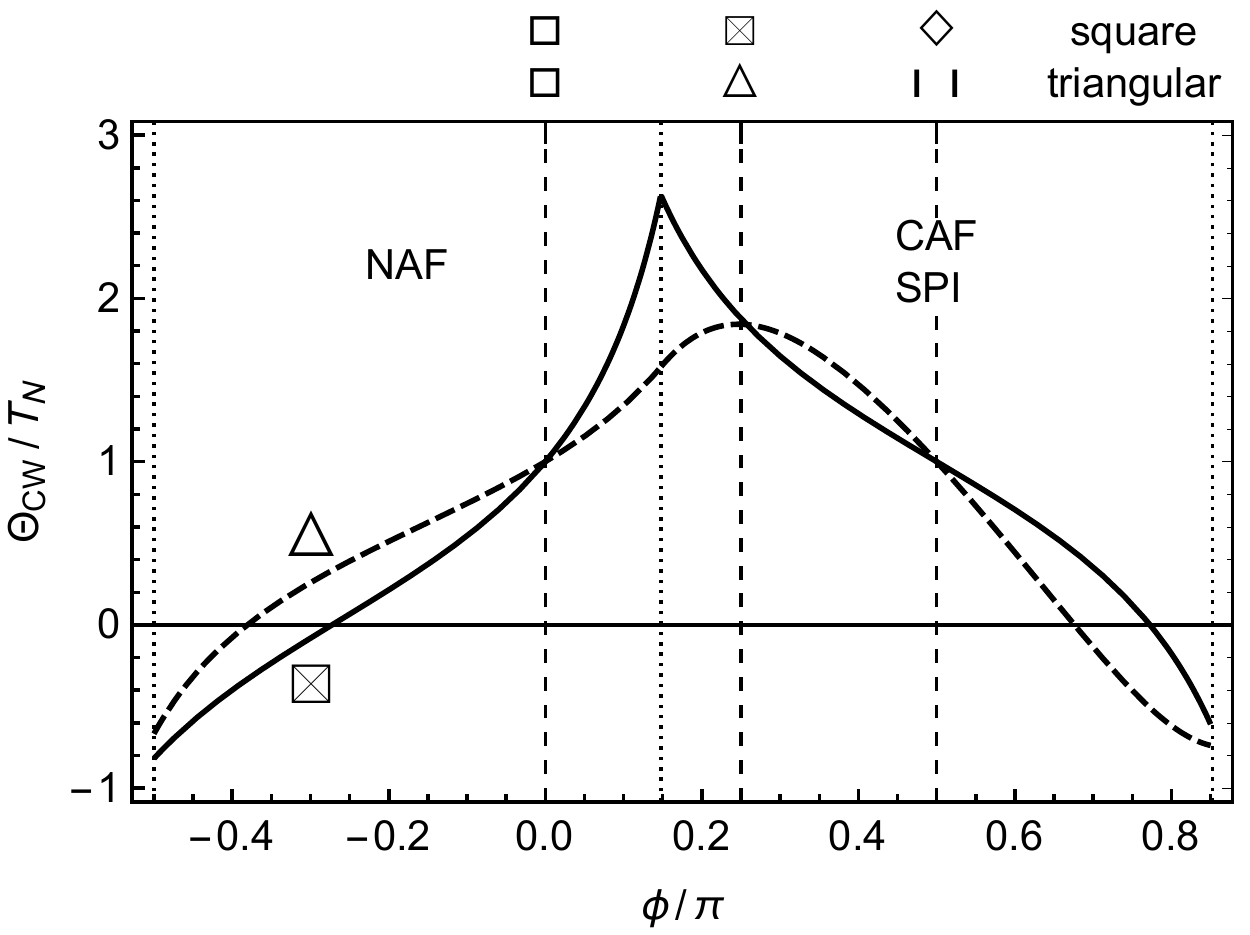}
	\caption{Frustration ratio $f=\Theta_\text{CW}/T_\text N$ using mean field $T_\text N$ for triangular (dashed line) 
	and square (solid line) lattice model with $J_\perp/J_\text c=0.2$. }
	\label{fig:trisqu_mf}
\end{figure}
It is not a priori obvious whether this widely used empirical criterion is sensible from a more microscopic point of view. An immediate problem  with the definition of $f$ is that the Curie-Weiss temperature, in particular, in frustrated magnets might be arbitrarily small as well. It can be even zero or negative, also for materials with AF order, due to competing interactions with opposite sign.
In the simple mean field approach applicable only for reasonably large $J_\perp/J_\text c$  the corresponding $f$ is shown in Fig.~\ref{fig:trisqu_mf}. Moderately enhanced values $|f|>1$ are only found around the NAF/CAF ($\square$) or NAF/SPI ($\triangle$) boundaries ($\phi/\pi=0.15$). On approaching the FM region from both sides $|f|$ does not show an enhancement due to the smallness of $|\Theta_\text{CW}|$. In fact for $\phi<0$ ($J_2<0$) this should not be expected because in this case the model is unfrustrated (Fig.~\ref{fig:frustration}).

On the other hand microscopically frustration is understood as the impossibility to minimize the ground state energy simultaneously for all exchange bonds. Therefore it appears natural to compare the ground state energy of the minimal building blocks of the frustrated lattice to the total ground state energy of the unfrustrated components. For example this can be achieved by defining the degree of frustration in the triangular lattice according to~\cite{schmidt:15}
\begin{equation}
\kappa_\triangle := 1-\frac{E_\triangle}{E_\text t+E_\text d}
\label{eqn:neel:kappa}
\end{equation}
for the frustrated triangle where $E_\triangle$ is its ground state energy and $E_{t,d}$ are those of its constituents, i.\,e. decoupled trimer and dimer. Explicitly~\cite{schmidt:15}
\begin{equation}
	E_\triangle(\phi)=\min
	\left(
		-\frac{3}{4}J_2,
		-J_1 +\frac14J_2,
		\frac12J_1+\frac14J_2
	\right)
\end{equation}
from which we also obtain $E_\text t :=E_\triangle(J_2=0)$ and $E_\text d :=E_\triangle(J_1=0)$. 
 A similar definition can be made for $\kappa_\boxtimes$  of the $J_1$-$J_2$ square lattice where the constituents are the unfrustrated square and the two diagonal dimer bonds. It turns out that $\kappa_\boxtimes(\phi)\equiv\kappa_\triangle(\phi)=:\kappa(\phi)$. 
This function indeed vanishes in the unfrustrated regime $J_2 <0$ or $-\pi\leq\phi\leq 0$. For $J_2>0$, the triangular lattice becomes frustrated, and $\kappa$ strictly monotonously increases until its maximum value $\kappa(\pi/4)=4/7\approx0.57$ which is the 2D isotropic point in the triangular phase diagram. Then $\kappa$ decreases to $\kappa(\pi/2)=0$ which is the point where the triangular lattice decouples into independent, unfrustrated AF chains. In the square lattice this case corresponds to two decoupled unfrustrated pure N\'eel sublattices. For ferromagnetic $J_1<0$ or $\phi>\pi/2$, $\kappa$ increases again to reach a maximum at the border between the spiral and FM phases. Therefore $\kappa$ peaks at or close to the strongly frustrated regions of the classical phase boundaries (Fig.~\ref{fig:frustration}) where magnetic order disappears. The large frustration is not only reflected in this ground state measure but also leads to signatures in the excited state spectrum. Full diagonalisation of small clusters~\cite{schmidt:17} shows that for $\phi$ values where $\kappa(\phi)$ approaches maximum the excited states are closely spaced and have large degeneracies. In the thermodynamic limit this signifies the strong suppression and breakdown of the ordered moment.

The  qualitative behavior of  $\kappa(\phi)$ therefore faithfully maps the degree of frustration as function of frustration control parameter $\phi$.
It is now a legitimate question to ask whether the quasi-2D Néel temperature $T_\text N(\phi)$ and empirical frustration ratio $|f(\phi)|$ show a
depression or enhancement, respectively, in the same region where $\kappa(\phi)$ is large.
For the simple mean field model the results are shown in Fig.~\ref{fig:trisqu_mf}. In fact at the $\phi/\pi\simeq 0.15$ phase boundary the peak in the frustration degree $\kappa(\phi)$ (Fig.~\ref{fig:frustration}) coincides with the enhancement of $f$. On the  $\phi/\pi\simeq 0.85$ classical boundary to the FM where $\Theta_\text{CW}$ has to change sign, however, no such coincidence is possible. It is important to investigate this further for the really interesting quasi-2D magnets. This requires a more advanced self-consistent RPA spin wave approach to calculate $T_\text N(\phi)$.

\section{LSW and RPA calculation of the quasi-2D N\'eel temperature}

A calculation of $T_\text N$ implies a theory of spin excitations at finite temperatures. This is a difficult problem from a fundamental point of view. In the linear spin wave (LSW) approximation and its various generalizations the spin excitations are described by bosons whose density increases with temperature, necessitating the inclusion of  interaction effects~\cite{zhitomirsky:13} beyond LSW. An effective empirical way to circumvent this difficult to treat many-body problem is provided by the Tyablikov method~\cite{tyablikov:67,irkhin:97} which assumes that the spin wave energy scale is reduced in accordance with  the decreasing ordered moment, instead of staying fixed as in LSW. It corresponds to an effective RPA approximation of the spin wave propagator~\cite{tyablikov:67,majlis:92,majlis:93}.  As noted in Ref.~\onlinecite{irkhin:97} the Tyablikov approach is `satisfactory from the practical but not from the theoretical point of view'. Since we take the former view and want to apply it to a general and practical understanding of frustration dependence of $T_\text N$ we use the Tyablikov approach, generalized to finite fields in this work. This is supported by a comparison to unbiased numerical MC simulations for the n.n. 2D Heisenberg model~\cite{yasuda:05} without field which prove that the numerical results show excellent agreement with RPA approximation throughout the  whole range of $J_\perp /J_1$ ratios.
\begin{figure}
	\includegraphics[width=.9\columnwidth]{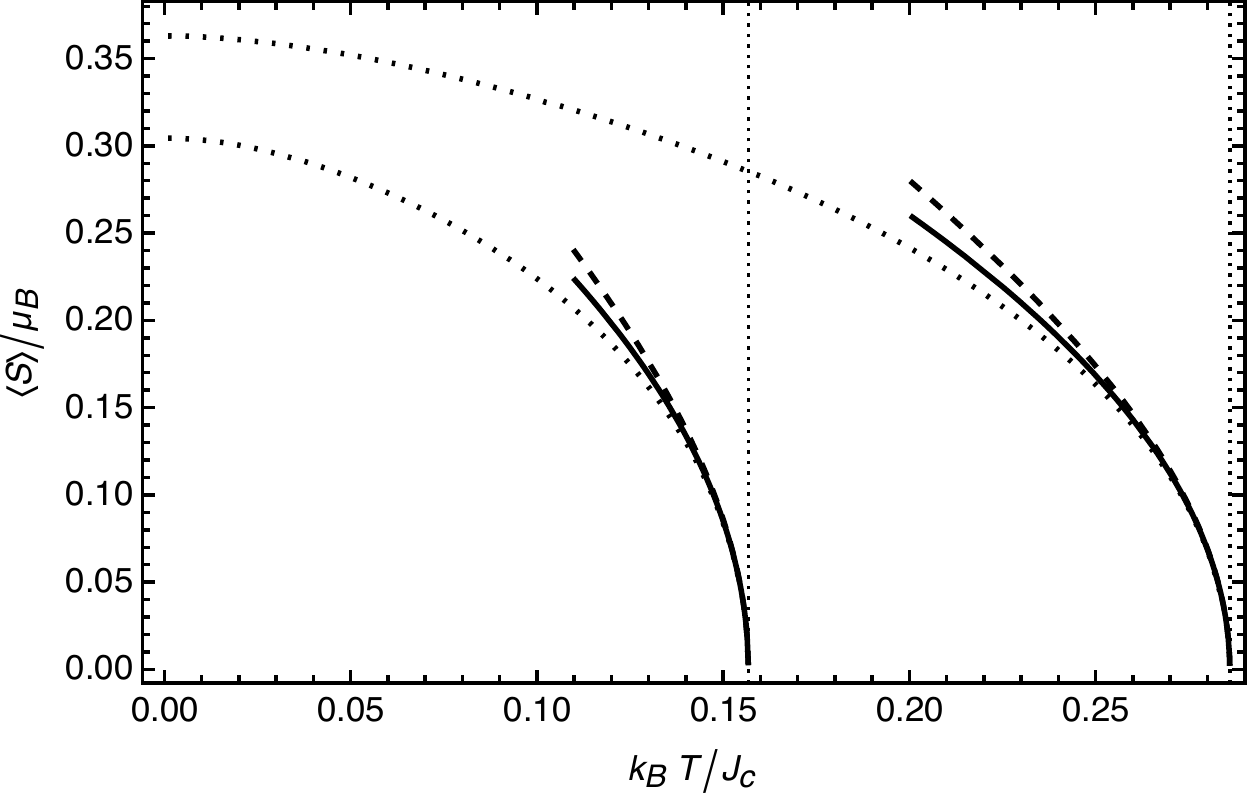}
	\caption{Ordered moment temperature dependence of square lattice for two frustration parameters $\phi\approx0$ (pure NAF, lower set of curves) 
	and $\phi\approx0.6\pi$ (CAF with FM $J_1$). The numerical solutions of Eq.~(\ref{eqn:neel:sbar2}) (dotted) are compared with analytical 
	solution of Eq.~(\ref{eqn:neel:sbarmf3}) comprising first (full line) and second (dashed line) order terms. The vertical dotted lines indicate the respective $T_\text N$.}
	\label{fig:msapprox}
\end{figure}

For the sake of a self-contained presentation we first recapitulate the LSW results of the $J_1$-$J_2$ model in an external field in a form that is applicable to square as well as triangular  lattice models. It is obtained from Eq.~(\ref{eqn:hex12}) by adding the Zeeman term leading to the full Hamiltonian
${\cal H}=  {\cal H}_\text{ex}-\vec h \cdot \sum_i\vec S_i$ with the definition $\vec h=g\mu_\text B\mu_0\vec H$.
In the LSW a Holstein-Primakoff (HP) transformation from local spin variables $S^\alpha_i$ ($\alpha=x,y,z$) to bosonic variables $a_i,a_i^\dagger$  at site $i$ is carried out using $S_i^+\rightarrow\sqrt{S/2}a_i$, $S_i^-\rightarrow \sqrt{S/2}a_i^\dagger$,  and  $S_i^z\rightarrow S-a_i^\dagger a_i$. Then, performing the Fourier transform 
\begin{equation}
a_i=\frac1{\sqrt N}\sum_{\vec k}a_{\vec k}^\dagger{\rm e}^{-{\rm i}\vec k\cdot\vec R_i}
\end{equation}
 the total Hamiltonian ${\cal H}$ may be written as a bilinear (harmonic) form in $\hat{a}^\dagger_{\vec k}=\left(a^\dagger_{\vec k}, a_{-\vec k}\right)$ which may be diagonalized (Eq.~(\ref{eqn:hdiagonal})) by the Bogoliubov transformation $\alpha_{\vec k}=u_{\vec k}a_{\vec k}+v_{\vec k}a_{-\vec k}^{\dagger}$
  to the magnon creation and annihilation operators $ \alpha_{\vec k},  \alpha_{\vec k}^\dagger$ of spin wave modes given in Eq.~(\ref{eqn:ek}). The transformation coefficients are obtained as
\begin{equation}
 \left\{
      \begin{array}{l}
      u_{\vec k}^2 \\   
      v_{\vec k}^2 \\
      \end{array}
    \right\}    
    =
    \frac12\left[
    \frac{A_{\vec k}
    -B_{\vec k}\cos^{2}\Theta_\text{cl}}
    {E_{\vec k}}
    \pm1\right]
    \label{eqn:bogol}
\end{equation}
with the sign convention $u_{\vec k}=\mathop{\rm sign}\nolimits B_{\vec k}|u_{\vec k}|$, $v_{\vec k}=|v_{\vec k}|$ and $SE_{\vec k}$ denoting the symmetric part of magnon energies given below. In the spirit of a (1/S) expansion the classical value $\Theta_\text{cl}$ of the moment canting angle ($\Theta_\text{cl}=\pi/2$ for zero field) is used in Eq.~(\ref{eqn:bogol}) as given by $\cos\Theta_\text{cl} =h/h_\text s$. Here $h_\text s=2SA_0=2S(J_0-J_{\vec Q})$ is the saturation field where the moments are aligned with the field  ($\Theta_\text{cl} =0$). This means $h_\text s/2S=4J_1$ for NAF and $h_\text s/2S=2(J_1+2J_2)$ for CAF. The final result of the HP transformation  is then the free magnon Hamiltonian
\begin{eqnarray}
{\cal H}&=&E_\text{cl}+E_\text{zp}+S\sum_{\vec k} E^\text{sw}_{\vec k}\alpha^\dagger_{\vec k}\alpha_{\vec k},\nonumber\\
E_\text{cl}&=&NS^2\left(J_{\vec Q}-A_0\cos^2\Theta_\text{cl}\right),\nonumber\\
E_\text{zp}&=&NSJ_{\vec Q}+\frac{S}{2}\sum_{\vec k}E_{\vec k}.
\label{eqn:hdiagonal}
\end{eqnarray}
Here $E_\text{cl}$ is the (negative) classical ground state energy, the second term $E_\text{zp}$ is the (negative) energy of zero point fluctuations of magnon modes and the last term describes the free Hamiltonian of excited magnons. The total ground state energy is  $E_\text{gs}=E_\text{cl}+E_\text{zp}$.  The zero point contribution is of relative order $1/S$ as compared to the classical part. The bare spin-wave or magnon energy  $\omega_{\vec k}=SE^\text{sw}_{\vec k}$ is obtained from the Bogoliubov transformation as 
\begin{eqnarray}
E^\text{sw}_{\vec k}&=&E_{\vec k}+E^\text a_{\vec k},
\nonumber\\
E_{\vec k}&=&
	\sqrt{
	\left[
	A_{\vec k}-B_{\vec k}
	\right]
	\left[
	A_{\vec k}+B_{\vec k}
	\left(1-2\cos^2\Theta_\text{cl}\right)
	\right],
    }
    \label{eqn:ek}
	\\
	E^\text a_{\vec k}&=& C_{\vec k}\cos\Theta_\text{cl},
	\nonumber
\end{eqnarray}
where intra- and inter-sublattice interactions $A_{\vec k}$ and  $B_{\vec k}$ as well as the interaction term $C_{\vec k}=-C_{-\vec k}$ which are symmetric and antisymmetric in $\vec k$, respectively are given by
\begin{eqnarray}
A_{\vec k}&=&J_{\vec k}
    +\frac12
    \left( 
    J_{\vec k+\vec Q}+J_{\vec k-\vec Q}
    \right)
    -2J_{\vec Q},
    \nonumber\\
     B_{\vec k}&=&J_{\vec k}
    -\frac12
    \left(
    J_{\vec k+\vec Q}+J_{\vec k-\vec Q}\right),
    \label{eqn:bk}
    \nonumber\\
    C_{\vec k}&=&J_{\vec k+\vec Q}-J_{\vec k-\vec Q}.
\label{eqn:swcoeff}    
\end{eqnarray}
\begin{figure}
\includegraphics[width=.9\columnwidth]{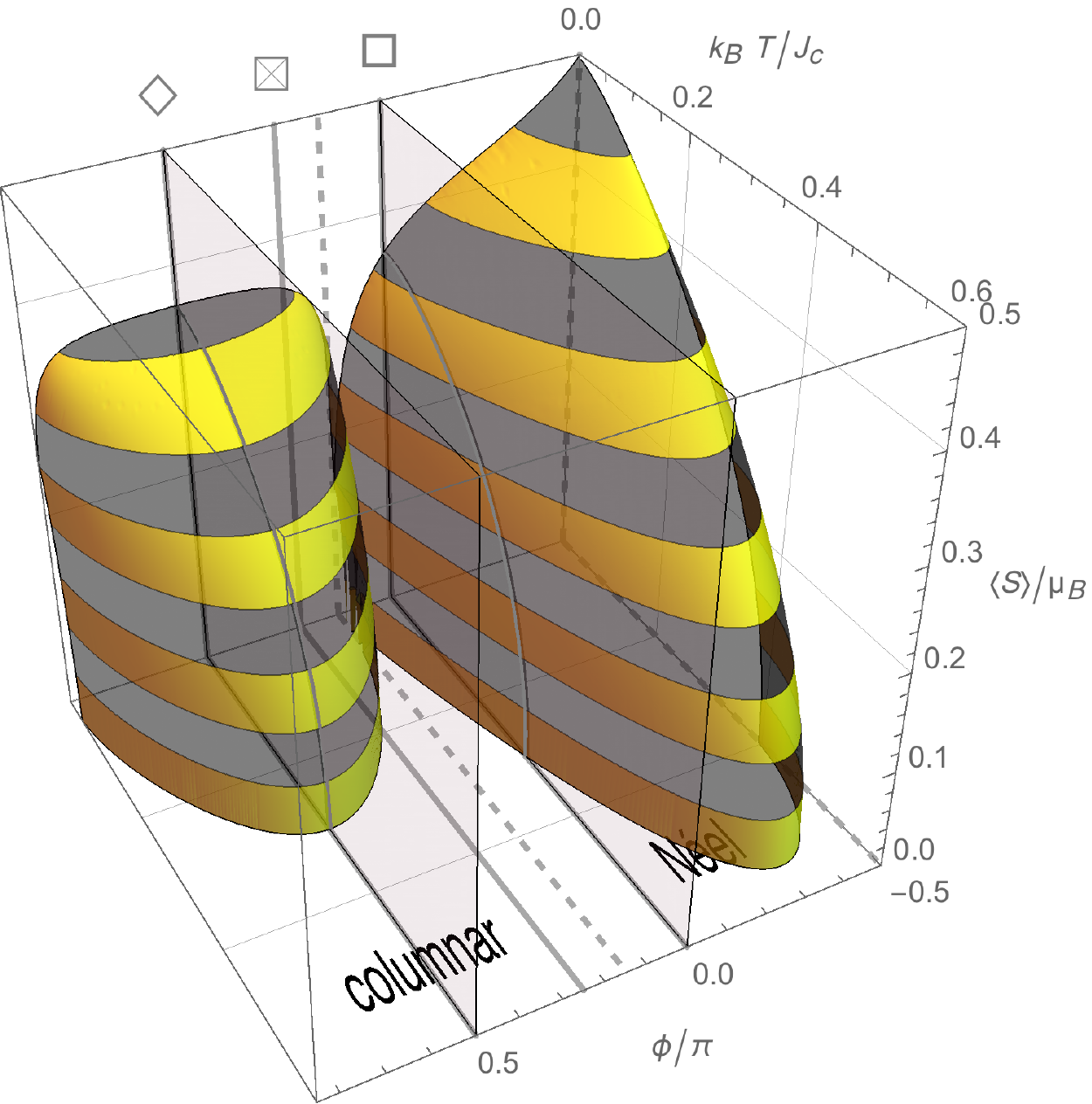}
\includegraphics[width=.9\columnwidth]{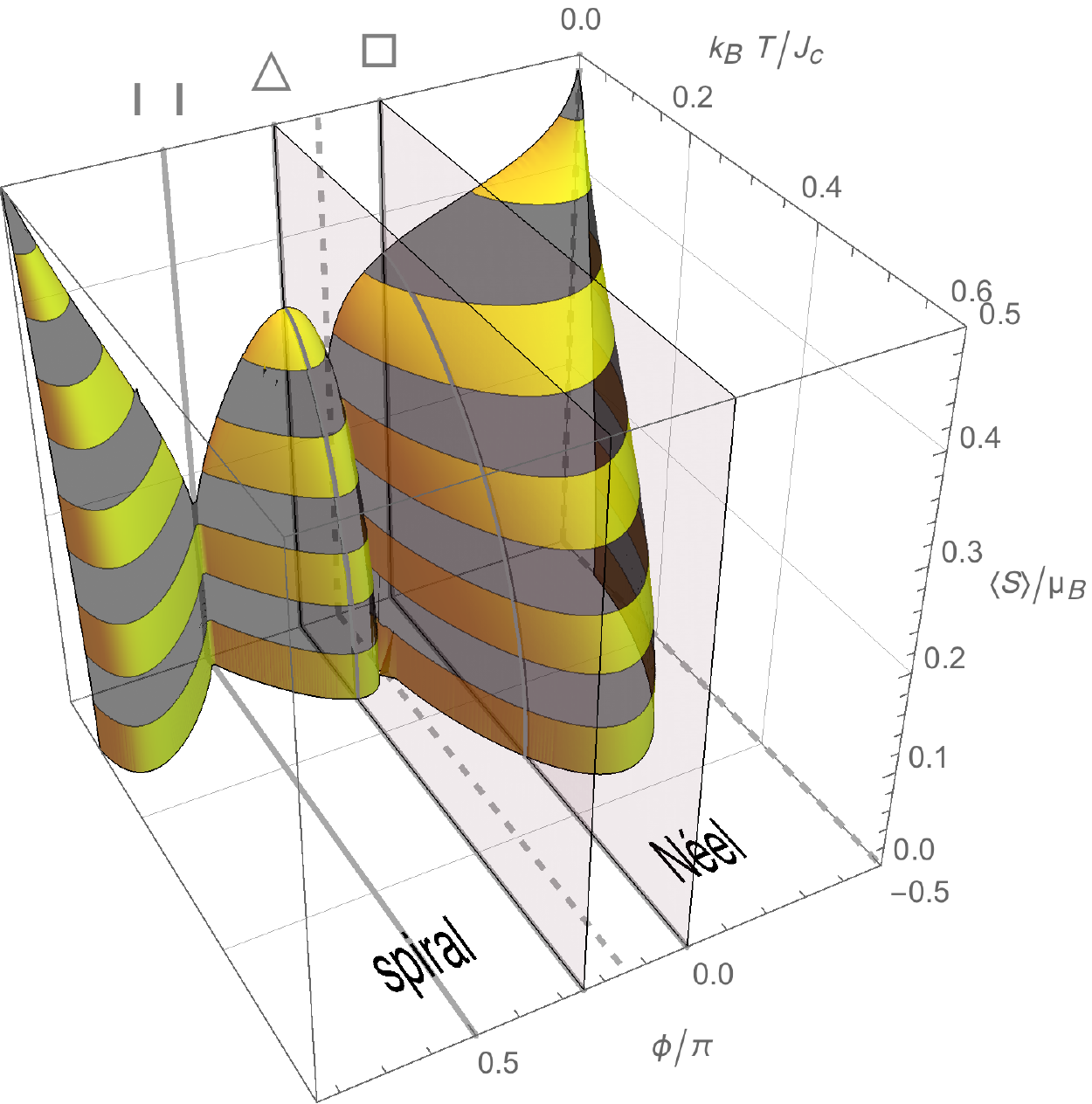}
\caption{Zero-field ordered moment $\langle S \rangle$ (ordinate) in  $\phi-T$ plane for the square lattice (a) and triangular lattice (b),
using $J_\perp/J_\text c=0.01$.
The $\langle S\rangle =0$ cut (basal plane) gives the $T_\text N$($\phi$) curves in Fig.~\ref{fig:square_TNf}a and 
Fig.~\ref{fig:triangular_TNf}a for square and triangular cases.
The $T=0$ cut (backside plane) likewise gives the ground-state ordered moment $m_{\vec Q}(\phi)$~\cite{schmidt:10, schmidt:11,schmidt:14}. 
Both curves are topologically equivalent and 
reach zero for $\phi$ values with strong frustration $(\phi/\pi\simeq 0.15,0.85)$. In (b) the additional zero at $\phi/\pi=0.5$ is 
not due to frustration but corresponds to unfrustrated quasi-1D AF chain case.}
\label{fig:sqtr_3Dms}
\end{figure}
Note that in general $E^\text{sw}_{\vec k}$ is {\em not\/} symmetric under $\vec k\rightarrow -\vec k$, 
because  $E^\text{sw}_{-\vec k}=E_{\vec k}-E^\text a_{\vec k}$ and therefore $E^\text{sw}_{\vec k}-E^\text{sw}_{-\vec k}=2E^\text a_{\vec k}\neq 0$. 
The asymmetric term is only relevant when $2\vec Q$ is {\em not\/} a reciprocal lattice vector, i.\,e. in the present
context only in the spiral phase of the triangular lattice for $\vec H\neq 0$.
In zero field where $E^\text a_{\vec k}=0$, $\Theta_\text{cl}=\pi/2$,   Eq.~(\ref{eqn:ek}) reduces to 
\begin{equation}
E^\text{sw}_{\vec k}=E_{\vec k}=
\sqrt{A^2_{\vec k}-B^2_{\vec k}}.
\label{eqn:ek0}
\end{equation}
Eqs.~(\ref{eqn:ek}) and (\ref{eqn:ek0}) are the basic quantities needed to calculate the N\'eel temperature and H-T phase diagram of the frustrated models within the Tyablikov RPA approach. This amounts to a stark simplification of the real interacting magnon problem. In fact due to the intrinsic interaction of magnons originating from higher order terms of the HP transformation already a zero temperature the magnon spectral function is renormalized~\cite{chernyshev:09}. In the strongly frustrated spin (dimer) liquid regime the spectrum may change qualitatively, consisting of a singlet bound state with finite gap, split-off from the two-magnon (triplon) continuum~\cite{kotov:99}. The gap closes at the quantum transition (as function of $\phi$) to the neighboring antiferromagnetic phases. These approaches are, however, difficult to generalize to finite temperature and finite interlayer coupling. As pointed out previously~\cite{irkhin:97} the empirical Tyablikov RPA method radically simplifies the problem by neglecting the change of spectral shape in the spin excitations due to multi-magnon interactions. It circumvents the complicated many-body processes by assuming that one still has a free magnon spectrum at higher temperature but with an overall dispersive width proportional to the $T$-dependent order parameter. This enforces a self-consistency condition from which $T_\text N(\phi,J_\perp)$ may be derived. While this seems acceptable in the magnetically ordered regimes it can only provide a qualitative interpolation across the narrow spin liquid regimes of $\phi$. In reality $T_\text N(\phi,J_\perp)$ has to be expected to be suppressed even more rapidly than predicted by the spin wave theory in these narrow $\phi$-intervals.

Rather than deriving the dynamical Green's function as in Ref.~\onlinecite{majlis:92} for the present purpose it is sufficient to calculate the self-consistent static moment directly. The condition that it vanishes will then determine the  N\'eel temperature. The total moment at a site $i$ is given by the thermal expectation value (with respect to $\cal{H}$)  $\left\langle S_i^{z'}\right\rangle$ in the {\em local\/} coordinate system the components of which we denote with a prime. In this coordinate system, the $z'$ axis at a given site site $i$ is aligned with the moment direction at that site. The latter is canted out of the plane given by the ordering vector $\vec Q$ due to the effect of the magnetic field which is directed along the global $z$ axis. The relation between moments in the local and global coordinate systems are given in Appendix~\ref{sect:app1}. 

In a finite magnetic field we have to distinguish three types of moments: The total moment $\langle S\rangle$, homogeneous moment $m_0$ and ordered moment $m_{\vec Q}$. While we consider all phases for zero field, in finite field  we will restrict to the commensurate CAF and NAF structures. For these coplanar cases the canted moments may be considered to lie in the $xz$ plane. Then we have the definitions
\begin{eqnarray}
\langle S\rangle
& =&
\frac{1}{N}\sum_i\left\langle S_i^{z'}\right\rangle, \nonumber\\
m_0& =&\frac{1}{N}\sum_i\left\langle S_i^z\right\rangle, \nonumber\\
m_{\vec Q}& =&\frac{1}{N}\sum_i{\rm e}^{{\rm i}\vec Q\cdot\vec R_i}\left\langle S_i^x\right\rangle.
\end{eqnarray}
(All moments are expressed in units of the Bohr magneton $\mu_\text B$.) Using the transformation in Appendix~\ref{sect:app1} one can verify that $m_0=\langle S\rangle\cos\Theta_\text{cl}$,  $m_{\vec Q}=\langle S\rangle\sin\Theta_\text{cl}$ and then
$\langle S\rangle^2 =m_0^2+m_{\vec Q}^2$.

For  $S=1/2$ we can write $S_i^{z'}=1/2-S_i^{-'}S_i^{+'}$. According to the linearized HP approximation for $S=1/2$ we then have $\langle S\rangle=(1/2)-2S\langle a_i^\dagger a_i\rangle$. In the moment reduction part of the right-hand side we now replace $S\rightarrow \langle S\rangle$. Physically this means that the reduction of the moment from $1/2$ due to the number of thermally excited Holstein-Primakoff bosons $\psi:=\left\langle a_i^\dagger a_i\right\rangle$ per site is rescaled by the already reduced moment $\langle S\rangle$. this substitution leads then to a self-consistency condition for the moment according to
\begin{equation}
	\langle S\rangle
	=
	\frac{1/2}{1+2\psi},
	\quad
	\psi
	=
	\left\langle a_i^\dagger a_i\right\rangle
	=
	\int_\text{BZ}\frac{{\rm d}^3k}{V_\text{BZ}}
	\left\langle a_{\vec k}^\dagger a_{\vec k}\right\rangle
	\label{eqn:selfcons1}
\end{equation}
where $\langle\cdots\rangle$ denotes the thermal average with respect to the magnon Hamiltonian of Eq.~(\ref{eqn:hdiagonal}). Unless otherwise noted, here and in the following we use continuum notation in reciprocal space, and integrations are done over the chemical Brillouin zone with volume $V_\text{BZ}$.

If $\psi$ is small (at low temperature) then $\langle S\rangle \simeq 1/2-\psi$  which recovers the LSW expression for the moment. The result of this simple physical consideration in Eq.~(\ref{eqn:selfcons1}) is equivalent to the RPA result~\cite{majlis:92} which also determines the temperature dependence of correlation functions in addition to the total moment. Here we only want to find the N\'eel temperature from the condition $m_{\vec Q}=0$. To this end one has to calculate $\psi=\psi(T,H)$ using the Bogoliubov transformation to magnon operators $\alpha_{\vec k},\alpha_{\vec k}^\dagger$ leading to
\begin{equation}
	\psi(T,H)
	=
	\int_\text{BZ}\frac{{\rm d}^3k}{V_\text{BZ}}
	\left[
	v_{\vec k}^2+\left(1+v_{\vec k}^2\right)
	\left\langle\alpha_{\vec k}^\dagger\alpha_{\vec k}\right\rangle
	+v_{\vec k}^2
	\left\langle\alpha_{-\vec k}^\dagger\alpha_{-\vec k}\right\rangle
	\right] 
\label{eqn:psidef1}
\end{equation}
Using Eq.~(\ref{eqn:bogol}) this may be evaluated to give the denominator in Eq.~(\ref{eqn:selfcons1}) and we finally obtain the self-consistency equation for the temperature- and field dependent total moment as
\begin{equation}
	\langle S\rangle
	=
	\frac12\left(
	\int_\text{BZ}\frac{{\rm d}^3k}{V_\text{BZ}}
	\frac{A_{\vec k}-B_{\vec k}\cos^2\Theta_\text{cl}}{E_{\vec k}}
	\coth\frac{\beta\langle S\rangle E^\text{sw}_{\vec k}}2
	\right)^{-1}
	\label{eqn:totalmoment}
\end{equation}
where $\beta=1/(k_\text BT)$. It is important to note that here $\Omega_{\vec k}=\langle S\rangle E^\text{sw}_{\vec k}$ is the modified magnon energy scaled by the temperature dependent prefactor $\langle S\rangle$ instead of the constant $S=1/2$ as in LSW approximation.

\begin{figure}
\includegraphics[width=.99\columnwidth]{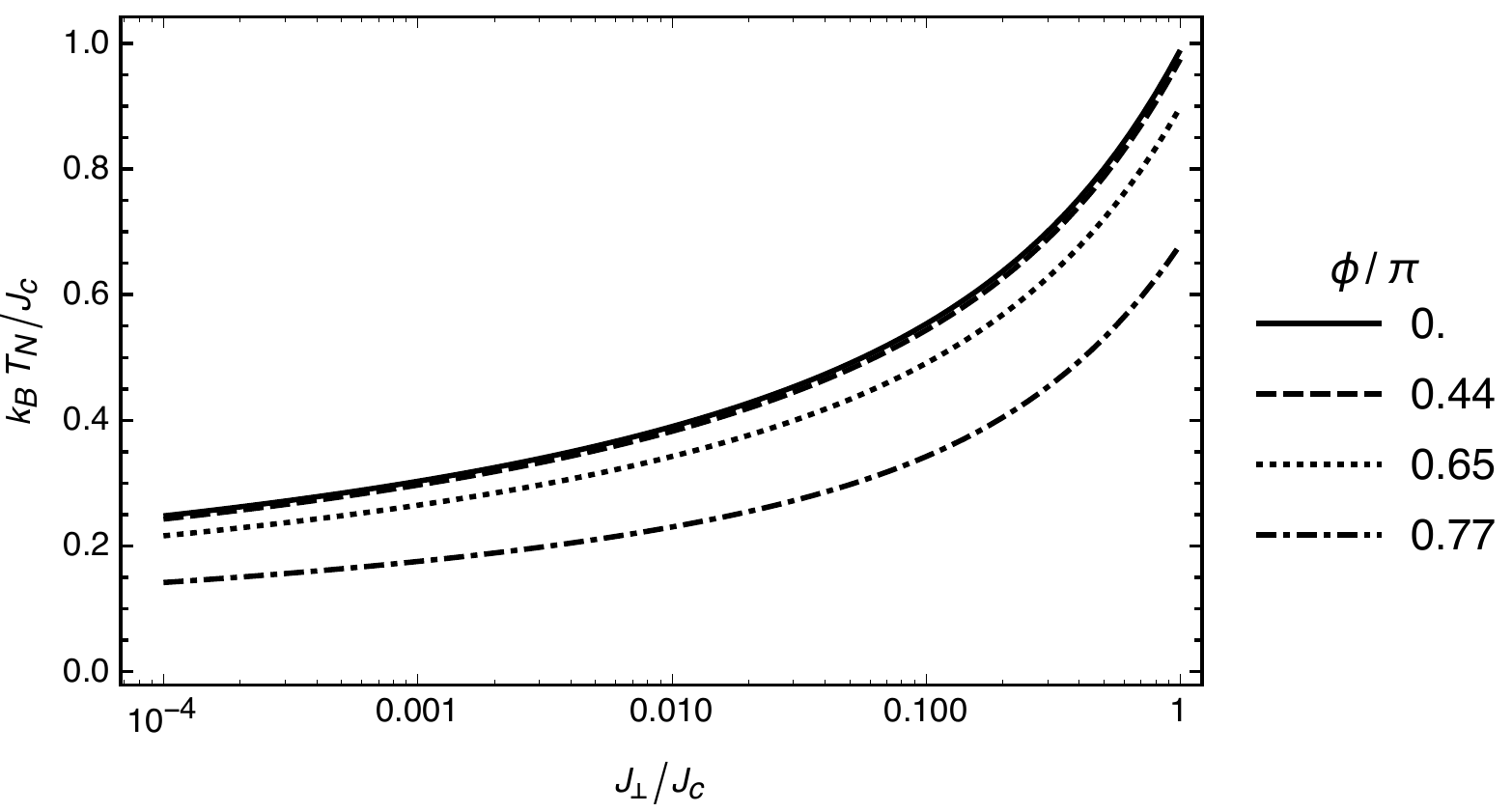}\hfill
\includegraphics[width=.795\columnwidth]{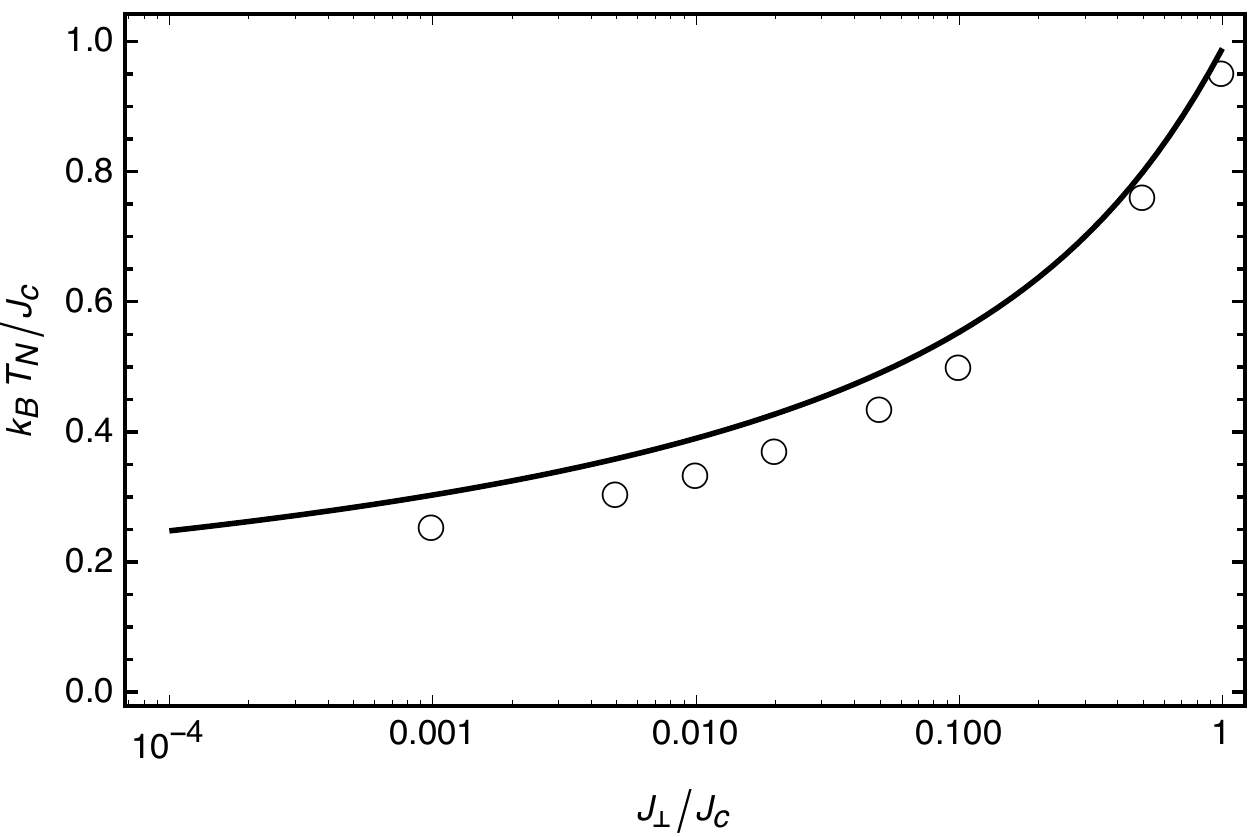}\hfill\null
\caption{(a) $T_\text N$ dependence on $J_\perp/J_\text c$ for various frustration control parameters $\phi$ in the square-lattice model in the NAF and CAF phases.
(b) $T_\text N$ for the unfrustrated ($\phi=0$ or $J_2=0$, $J_\text c=J_1$) pure N\'eel case. Full line same as in (a). The open circles are from MC simulations in Ref.~\onlinecite{yasuda:05}. Note the additional ordinate scale factor $S(S+1)=3/4$ employed in this reference.}
\label{fig:frustratio}
\end{figure}
The Néel temperature $T_\text N=1/(k_\text B\beta_\text N)$ itself is defined as the temperature where the ordered moment vanishes. For small magnetic fields with $h/h_\text s\ll\beta_\text N(h=0)J_\text c/4$, we can identify the ordered moment $m_{\vec Q}$ with the total moment $\langle S\rangle$. Expanding Eq.~(\ref{eqn:totalmoment}) for $\langle S\rangle\ll1$ to leading order, we obtain a closed expression
\begin{equation}
	\beta_\text N(h)
    =
    4\int_\text{BZ}\frac{{\rm d}^3k}{V_\text{BZ}}
    \frac{A_{\vec k}-B_{\vec k}\cos^{2}\Theta_\text{cl}}
    {E_{\vec k}^2-E_{\vec k}^{a2}}.
    \label{eqn:betan}
\end{equation}
Below $T_\text N$ the $T$ dependence of the ordered moment $\langle S\rangle$ is obtained by an iterative solution of Eq.~(\ref{eqn:totalmoment}). To improve numerical convergence it is preferable to separate out the singular term in the integrand by defining $g(x)=\coth(x)-1/x$ leading to a numerically more suitable form of the self-consistency equation:
\begin{eqnarray}
	\langle S\rangle
	&=&
	\frac12\left(1-\frac{\beta_\text N}\beta\right)\times
	\nonumber\\
	&&
	\left[
	\int_\text{BZ}\frac{{\rm d}^3k}{V_\text{BZ}}
	\frac{A_{\vec k}-B_{\vec k}\cos^2\Theta_\text{cl}}{E_{\vec k}}
	g\left(\frac{\beta\langle S\rangle E_{\vec k}}2\right)
	\right]^{-1}
	\label{eqn:neel:sbar2}
\end{eqnarray}
where the integral in brackets is now a well-behaved function near $T=T_\text N$. Its expansion for small $\langle S\rangle$ leads to an approximate expression close to $T_\text N$:
\begin{eqnarray}
	\langle S\rangle
	&\approx&
	\sqrt{\frac{1-\beta_\text N/\beta}{\beta_\text NI_0}}
	\left[
	1+\frac12\left(1-\frac{\beta_\text N}{\beta}\right)
	\frac{\beta_\text NI_1}{I_0^2}
	\right.
	\nonumber\\
	&&
	\left.
	-\frac12
	\left(1-\frac{\beta_\text N}{\beta}\right)^2
	\frac{\beta_\text N^2}{I_0^4}
	\left(I_0I_2-\frac{7}{4}I_1^2\right)
	\right]	
	\label{eqn:neel:sbarmf3}
\end{eqnarray}
where the expansion integrals $I_{0-2}$ are given in Appendix \ref{sect:app2}. An comparison of numerical solution and analytical approximation for a CAF and NAF case is shown in Fig.~(\ref{fig:msapprox}). The global behavior of the zero-field total moment $\langle S\rangle(T,\phi)$ as function of temperature and frustration control parameter is depicted in the 3D plot of Fig.~\ref{fig:sqtr_3Dms} for both lattices.

The self-consistency equation Eq.~(\ref{eqn:totalmoment}) may also be used to calculate the field dependence of the N\'eel temperature. It is defined by the condition that the order parameter vanishes, i.\,e. $m_{\vec Q}=0$. Then
the total moment is equal to the magnetization per site $\langle S\rangle=m_0$. For simplicity, we use the classical value $m_0=(1/2)\cos\Theta_\text{cl}$. Furthermore $m_0$ may be taken as $T$ independent as long as $k_\text BT_\text N\ll J_\text c$.

Replacing $\langle S\rangle\rightarrow m_0$ in Eq.~(\ref{eqn:totalmoment}) gives an implicit equation for $T_\text N=T_\text N(h)$.
It may be presented in a form more convenient for numerical solution as
\begin{eqnarray}
	T_\text N
	&=&
	\tilde T_\text N
	\left[\vphantom{\left(\frac{m_0E_{\vec k}}{2k_\text BT_\text N}\right)}
	1-2m_0\times
	\right.
	\nonumber\\&&
	\hphantom{\tilde T_\text N}
	\left.
	\int_\text{BZ}\frac{{\rm d}^3k}{V_\text{BZ}}
	\frac{A_{\vec k}-B_{\vec k}\cos^2\Theta_\text{cl}}{E_{\vec k}}
	g\left(\frac{m_0E_{\vec k}}{2k_\text BT_\text N}\right)
	\right]
	\label{eqn:TNH}
\end{eqnarray}
where $\tilde T_\text N=1/(k_\text B\beta_\text N)$ now denotes the small-field expression from Eq.~(\ref{eqn:betan}). For $h\ll h_\text s$ when $m_0\rightarrow 0$ we indeed recover Eq.~(\ref{eqn:betan}).

For $h=0$ and for general $h$ in the nonspiral (NAF, CAF) phases  the asymmetric term $\sim E^{a2}_{\vec k}$ in the spin wave dispersion vanishes and $E^\text{sw}_{\vec k}=E_{\vec k}$. 
The zero-field limit $T_\text N^0:=T_\text N(h\to0)$ is then given by~\cite{majlis:92}
\begin{equation}
	T^0_\text N
	=
	\left(4k_\text B
	\int_\text{BZ}\frac{{\rm d}^3k}{V_\text{BZ}}
	\frac{A_{\vec k}}{E^2_{\vec k}}
	\right)^{-1}.
	\label{eqn:TN0}
\end{equation}
This is an explicit expression for $T^0_\text N$ which does not contain $T^0_\text N$ on the r.h.s. any more. It properly reproduces the 2D limit $T^0_\text N\to0$ when the $k_z$- dispersion of magnons vanishes and the integral in Eq.~(\ref{eqn:TN0}) diverges logarithmically (see Sec.~\ref{sec:analytical}). The above equations contain all information on the frustration effect encoded in the spin wave expressions $A_{\vec k}, B_{\vec k}$ and $E_{\vec k}$ of Eq.~(\ref{eqn:ek},\ref{eqn:swcoeff}).

\section{Numerical results for the zero-field N\'eel temperature}
\label{sec:numerical}

In this section we discuss the systematic variation of $T_\text N(\phi)$ with frustration control parameter  $\phi=\tan^{-1}(J_2/J_1)$ as obtained from the numerical calculations based on Eq.~(\ref{eqn:TN0}) for square as well as triangular models.

\subsection{Square-lattice model}

Fig.~\ref{fig:frustratio}a displays the dependence of $T_\text N$ obtained from Eq.~(\ref{eqn:TN0}) on the interplane coupling $J_\perp$ for selected values of $\phi$.  The approximately logarithmic variation with $J_\perp/J_\text c$ known from the pure N\'eel case is observed to hold also in the frustrated case (see Sec.~\ref{sec:analytical}). In this case ($J_2=0$, $J_\text c=J_1$) there are MC simulation results~\cite{yasuda:05} for $T_N$ which can be compared to the results of the present theory (Fig.~\ref{fig:frustratio}b). They show a good agreement within only a few per cent deviation in the whole range of $J_\perp/J_\text c$ plotted.~\footnote{Note that in this work the $T_\text N$ curves are resulting from the condition of vanishing order parameter which is calculated in selfconsistent TA approach, whereas in Ref.~\onlinecite{yasuda:05} $T_\text N$ is obtained from the condition of the divergent RPA susceptibility of a quasi-2D N\'eel antiferromagnet ($J_2=0$) above $T_\text N$.}

\begin{figure}
	\includegraphics[width=\columnwidth]{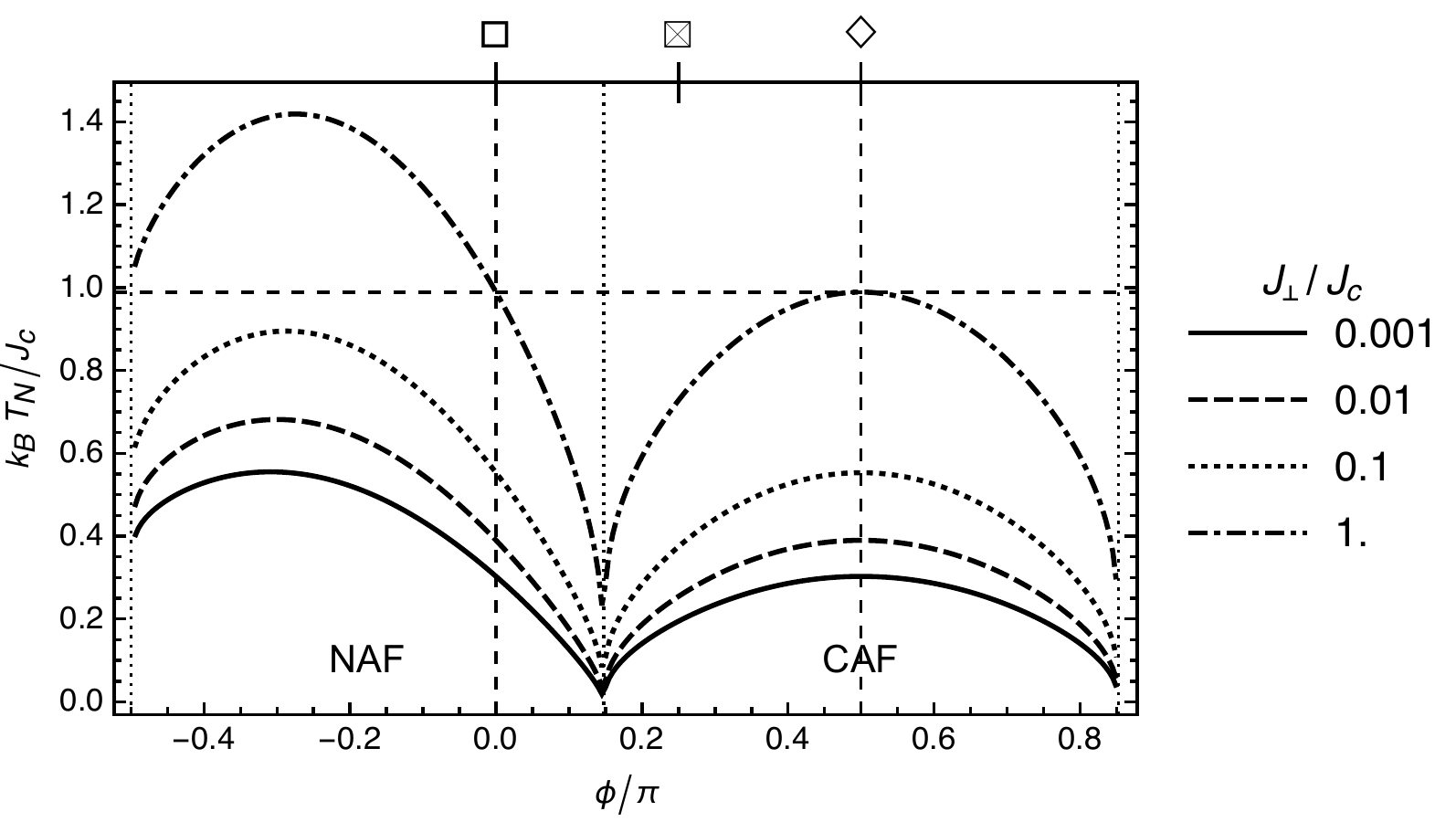}
	\includegraphics[width=\columnwidth]{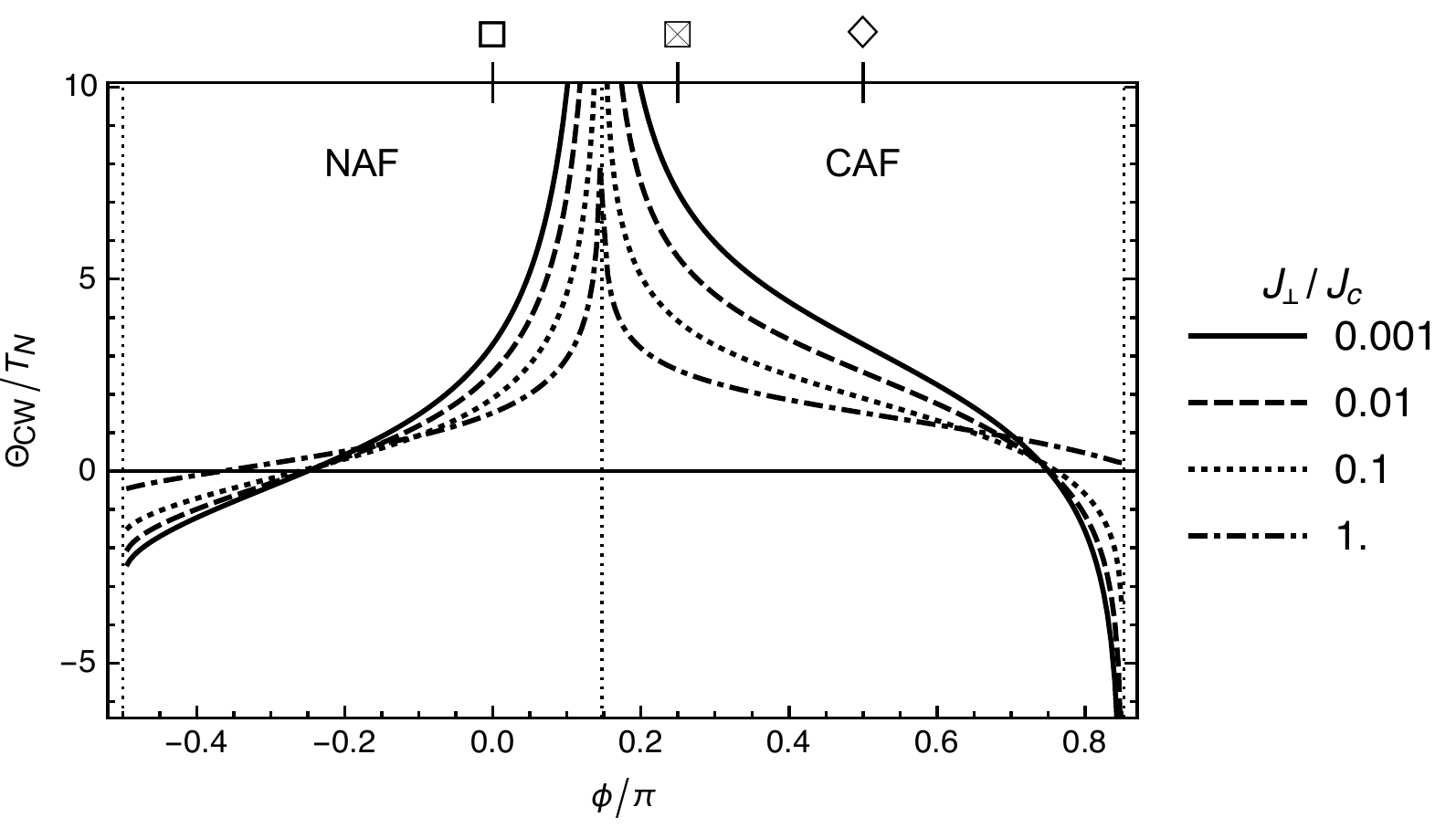}
	\caption{Quasi-2D N\'eel temperature $T_\text N$ (a) and empirical frustration ratio $f=\Theta_\text{CW}/T_\text N$ 
	(b) for the square-lattice $J_1$-$J_2$ as function of frustration control parameter $\phi$ for different interlayer 
	coupling strengths $J_\perp/J_\text c =10^{-3} ...1$.}
	\label{fig:square_TNf}
\end{figure}

In the complementary Fig.~\ref{fig:square_TNf}a we show the Néel temperature $T_\text N(\phi)$  at different strengths of the inter-plane coupling $J_\perp/J_\text c=10^{-3}\ldots1$. $T_\text N$ vanishes at the borders of the columnar phase, see Appendix~\ref{sec:tncafborders}. In the isotropic 3D cases with $J_1=J_\perp=J_\text c$, $J_2=0$, and $J_1=0$, $J_2=J_\perp=J_\text c$, the result $k_\text BT_\text N/J_\text c\approx0.989$ is obtained. Note that the corresponding MF result (Eq.~\ref{eqn:mftn}) would be $k_\text BT_\text N/J_\text c=3/2$.  The symmetric $\phi$ dependence in the CAF phase is due to a mirror symmetry of the Hamiltonian at $J_1=0$ ($\phi=\pm\pi/2$): The square lattice is bipartite and the Hamiltonian remains invariant upon a sign change of all spins on one sublattice while simultaneously replacing $J_1\to-J_1$. This transforms the Néel antiferromagnet to a ferromagnet (not shown) and the CAF phase with $(\pi,0)$ ordering into $(0,\pi)$ ordering.

For the $S=1/2$ frustrated square lattice, the Curie-Weiss temperature is given by $\Theta_\text{CW}=(J_1+J_2+J_\perp/2)/k_\text B$ (Eq.~(\ref{eqn:TCW})). Fig.~\ref{fig:square_TNf}b displays the corresponding parameter $f=\Theta_\text{CW}/T_\text N$ as a function of $\phi$ for different interlayer coupling strengths $J_\perp/J_\text c=10^{-3}\ldots1$. The overall $\phi$ dependence is in good agreement with the approximate analytical evaluation of Eq.~(\ref{eqn:TN0}) in Sec.~\ref{sec:analytical}.

It is instructive to compare $f(\phi)$ with the behavior of the microscopic frustration degree $\kappa(\phi)$ shown in Fig.~\ref{fig:frustration} (Eq.~(\ref{eqn:neel:kappa})). In the Néel phase, we indeed obtain a correspondence between $\kappa$ and $f$: Where $\kappa\equiv0$, in the whole unfrustrated NAF phase as well as at $\phi=\pi/2$, we correspondingly obtain $|f|={\cal O}(1)$. This is true even for the quasi-2D case and $|f|$ appears to increase only logarithmically with decreasing $J_\perp/J_\text c$. (In the regions $\phi<-\pi/4$ $f$ turns negative because $\Theta_\text{CW}<0$ due to a ferromagnetic $J_2<-J_1<0$.)

A finite positive $J_2$ turns on frustration with $\kappa>0$. In the Néel phase, this is reflected by a corresponding increase in $f$ which eventually diverges at the NAF/CAF border. Here, $J_2=J_1/2$, and $\kappa =4/11\approx0.36$.

The analogy between $\kappa$ and $f$ partially fails in the CAF phase: $\kappa$ does not show any special feature at the NAF/CAF border but instead increases strictly monotonously to its maximum value $\kappa =4/7\approx0.57$ at $J_2=J_1$ ($\phi=\pi/4$). It then decreases and vanishes again at the special point $\phi=\pi/2$ where $J_1=0$ with decoupled sublattices that correspond to the pure N\'eel case of $\phi=0$. For $\phi>\pi/2$, $J_1$ turns ferromagnetic, leading to an increase in $\kappa $ until the FM border of the CAF phase where $J_2=-J_1/2$ and $\kappa=4/7$ again. Different to the behavior of $\kappa$, $|f|$ is diverging at both the border to the NAF and the border to the FM phase. With increasing $\phi$, $f$ decreases strictly monotonously, crossing $f=0$ at $\phi>3\pi/4$ due to the sign change of $\Theta_\text{CW}$.

\subsection{Anisotropic triangular-lattice model}

\begin{figure}
	\includegraphics[width=\columnwidth]{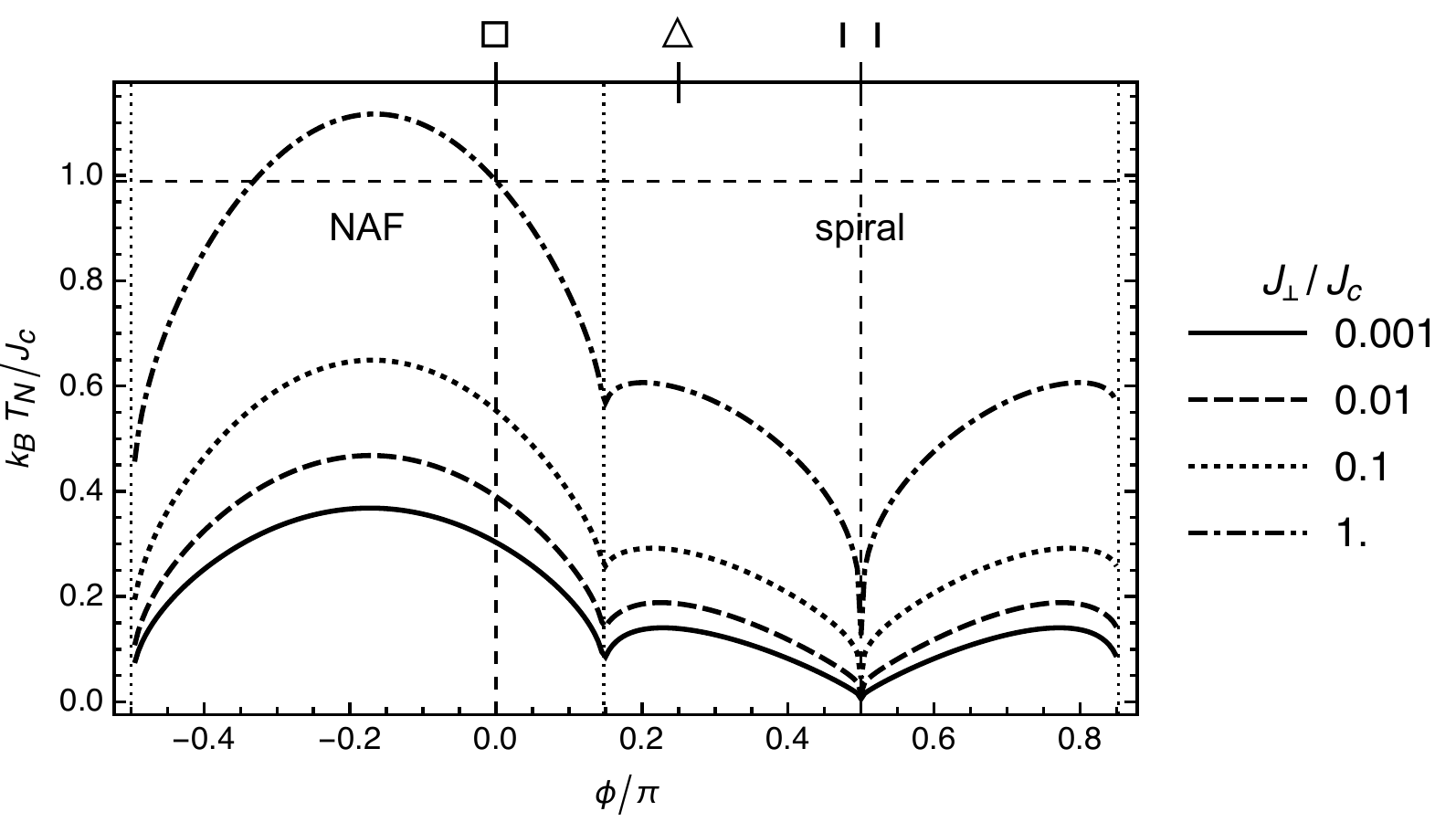}
	\includegraphics[width=\columnwidth]{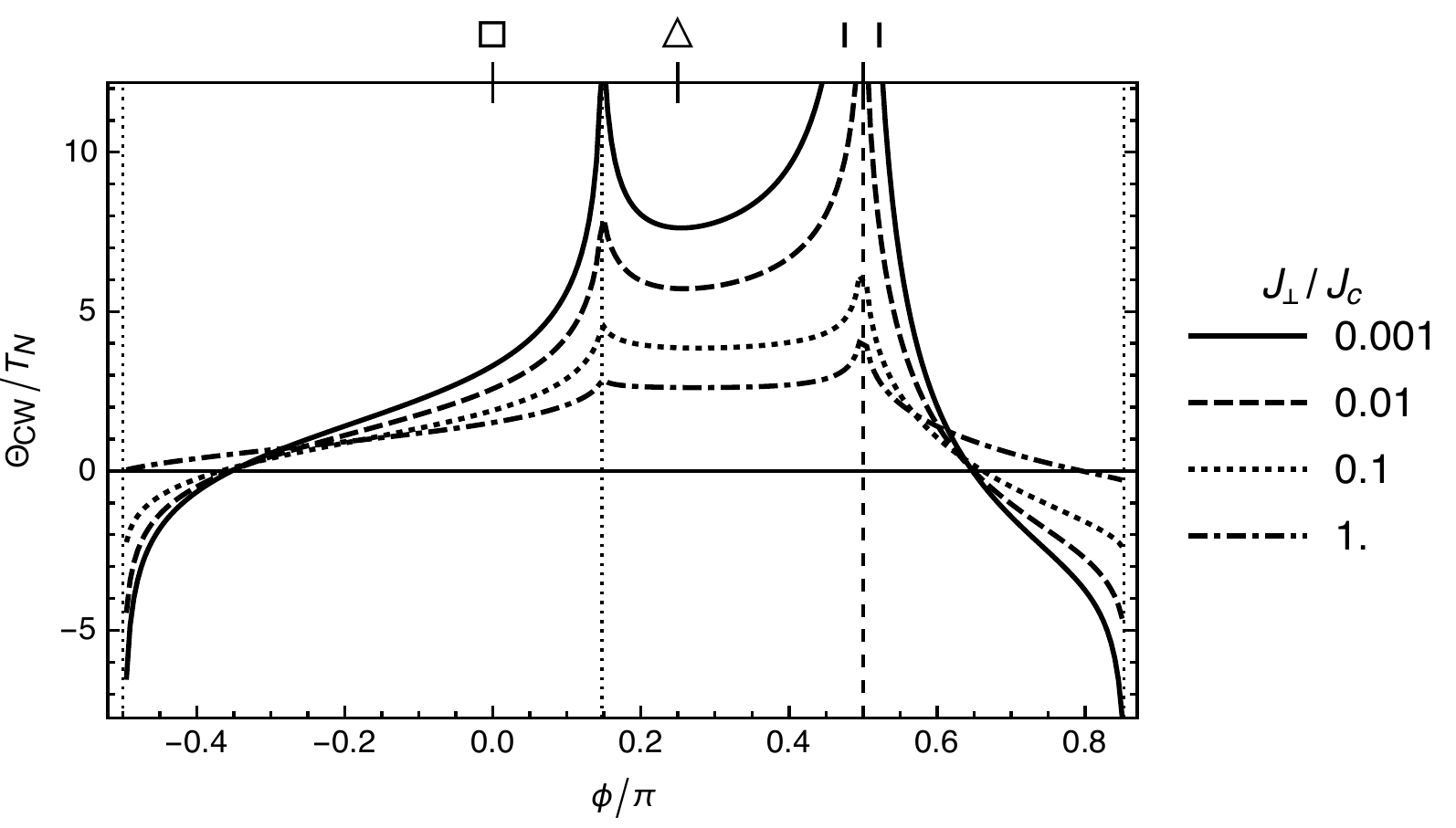}
	\caption{Quasi-2D N\'eel temperature $T_\text N$ (a) and empirical frustration ratio $f=\Theta_\text{CW}/T_\text N$
	 (b) for the triangular-lattice $J_1$-$J_2$ as function of frustration control parameter $\phi$ for different interlayer coupling 
	 strengths $J_\perp/J_\text c =10^{-3} ...1$. }
	\label{fig:triangular_TNf}
\end{figure}

In Fig.~\ref{fig:triangular_TNf}a we show the parameter dependence of the Néel temperature for the $S=1/2$ anisotropic triangular lattice. Different curves correspond to different interplane couplings $J_\perp/J_\text c=10^{-3}\ldots1$. In the Néel phase for $-\pi/2\le\phi\le\tan^{-1}(1/2)\approx0.15\pi$, the qualitative behavior is similar to the square-lattice case: $T_\text N$ increases from zero at $\phi=-\pi/2$ ($J_1=0$, $J_2=-J_\text c<0$) to a maximum in the middle of the NAF phase and decreases to a cusp-like minimum at $J_2/J_1=1/2$ ($\phi\approx0.15\pi$), the border with the spiral phase in the LSW approximation. The Néel temperature decreases again towards $T_\text N=0$ at $\phi=\pi/2$ ($J_1=0$, $J_2=J_\text c>0$). At this particular point in the phase diagram the triangular lattice turns into an unfrustrated set of independent one-dimensional AF $J_2$ chains coupled by $J_\perp$, i.\,e. a strictly 2D magnet. The ordering temperature thus vanishes {\em not\/} due to frustration but can be understood as a consequence of the Mermin-Wagner theorem. The mirror symmetry in the spiral phase around $\phi=\pi/2$ is due to the same mirror symmetry present in the Hamiltonian for the square lattice, regarding the anisotropic triangular lattice as a depleted $J_1$-$J_2$ square lattice where one set of diagonal $J_2$ bonds is missing.

It is again useful to compare $f(\phi)$  with the microscopic degree of frustration $\kappa(\phi)$  (Fig.~\ref{fig:frustration}). Now the Curie-Weiss temperature is given by $\Theta_\text{CW}=(J_1+J_2/2+J_\perp/2)/k_\text B$ (Eq.~(\ref{eqn:TCW})). In the Néel phase with ferromagnetic $J_2<0$ ($\kappa=0$, unfrustrated for $-\pi/2\le\phi\le0$), we generally obtain $|f|={\cal O}(1)$ for $J_\perp=J_\text c$ and its maximum value increases only logarithmically with decreasing interlayer coupling $0<J_\perp/J_\text c<1$. This is similar to the square-lattice case. However except for an isotropic interlayer coupling $J_\perp=J_\text c$ where $\Theta_\text{CW}=0$, $|f|$ diverges at the  NAF/FM border ($J_1=0$ or $\phi/\pi=-0.5$). Like for antiferromagnetic $J_2$ as discussed in the preceding paragraph, this divergence is caused by a vanishing $T_\text N$ at the border which is also due to the formation of a strictly two-dimensional system consisting of ferromagnetic $J_2$ chains coupled with $J_\perp$.

A peak appears at the NAF/SPI border $(\phi/\pi=0.15)$ which is, as in the square-lattice case, not reflected by any special feature in $\kappa(\phi)$. At $J_1=0$ or $\phi/\pi=0.5$ again a divergence due to vanishing $T_\text N$ appears, this time at the antiferromagnetic $J_2>0$ side of the phase diagram. We have $\kappa(\pi/2)=0$ here, because at this point, the model is unfrustrated and the divergence of $f$ is exclusively due to the previously discussed lowering of dimensionality. Another peak in $|f|$ for small $J_\perp$ is present at the SPI/FM boundary  ($J_2=-J_1/2$, $\phi/\pi\approx0.85$).

Altogether only the two peaks in $|f(\phi)|$ at the NAF/SPI and SPI/FM boundaries can be associated with the regions of high frustration (large $\kappa(\phi)$ in Fig.~\ref{fig:frustration}). The divergences in $|f|$ at $\phi=\pm\pi/2$ in contrast are unrelated to frustration effects but are due to dimensional reduction only. This does not appear in the square lattice (Fig.~\ref{fig:square_TNf}) because of the additional $J_2$ bond.

\section{Analytical approximations for the N\'eel temperature}
\label{sec:analytical}

It is useful to complement the numerical determination of zero-field $T_\text N(\phi,J_\perp)$ with approximate analytical results to gain a better understanding of the frustration influence. They are derived by expanding the integrand in Eq.~(\ref{eqn:TN0}) for small $\vec k$ vectors where $E_{\vec k}$ tends to zero; this region dominates the value of the integral. For NAF there are two equivalent dispersion minima $\vec k\approx (0,0), (\pi,\pi)$ for NAF phase and four equivalent minima positions $\vec k\approx (0,0), (0,\pi), (\pi,0), (\pi,\pi)$ for CAF phase. The expansion has to be done separately for each magnetic structure and the cutoff wave vectors in both cases have to be chosen such that the symmetry $T_\text N(J_1=0)=T_\text N(J_2=0)$ is preserved. We will restrict ourselves in this section to the NAF and CAF phases of the square lattice only.

\subsection{NAF structure ($2J_2 <J_1$)}
\label{sec:approxNAF}
Here the expansion of $E_{\vec k}$ and $A_{\vec k}$ to lowest order leads to 
\begin{equation}
k_\text BT_\text N=
\left[\frac{4}{\pi^3}
\int_0^\pi\int_0^{k_\parallel^\text c}\frac{d^2k_\parallel}{\tilde J k_\parallel^2+J_\perp k_z^2}
\right]^{-1}
\label{eqn:TNNAF1}
\end{equation}
where we defined the effective exchange  $\tilde J :=J_1-2J_2 >0$ for the frustrated NAF and  $\epsilon^2 :=J_\perp/\tilde J$ as the parameter that measures the relative strength of interlayer coupling. The integral diverges, i.\,e., $T_\text N\rightarrow 0$ in the purely 2D case ($J_\perp=0$) and also when approaching the classical NAF/CAF phase boundary at $\tilde J=0$ ($2J_2=J_1$) where the ordered moment vanishes due to strong frustration effect. The value of $T_\text N$ depends weakly on the cutoff which we choose as $k^\text c_\parallel =\pi$ corresponding to a Debye-approximation for the spin wave spectrum.. The evaluation of Eq.~(\ref{eqn:TNNAF1}) leads to
\begin{eqnarray}
	k_\text BT_\text N=(J_1-2J_2)\frac{\pi}{B_\epsilon
	+\ln\left(1+\frac1{\epsilon^2}\right)}.
\label{eqn:TNNAF2}
\end{eqnarray}
Here $B_\epsilon=(1/\epsilon)(\pi-2\tan^{-1}(1/\epsilon))$.  This expressions hold for the whole frustrated NAF region $-\pi/2<\phi<0.15\pi$.
It is useful to derive the approximate expression (except very close to $J_1=2J_2$) for the extreme quasi-2D case with $\epsilon^2\ll1$. We obtain
\begin{eqnarray}
k_\text BT_\text N \approx(J_1-2J_2)\frac{\pi}{2+\ln\left(\frac{J_1-2J_2}{J_\perp}\right)}.
\label{eqn:TNNAF3}
\end{eqnarray}
For $\epsilon^2<0.1$ this is indistinguishable from Eq.~(\ref{eqn:TNNAF2}).

We note that the natural exchange scale that determines $T_{N}$ is really $\tilde J$ rather than $J_\text c$. The sign of $\tilde J$ changes at the NAF/CAF boundary and in the CAF phase it simply has to be replaced with $|\tilde J|$. Using the (3D) ordering vectors  $\vec Q_\text{NAF}=(\pi,\pi,\pi)$ and $\vec Q_\text{CAF}=(\pi,0,\pi)$ and Eq.~(\ref{eqn:exfourier}) we may also express it as 
\begin{equation}
\tilde J=J_1-2J_2=\frac12\left(J_{\vec Q_\text{CAF}}-J_{\vec Q_\text{NAF}}\right).
\end{equation}
For the unfrustrated ($J_2=0$) NAF Eq.~(\ref{eqn:TNNAF2}) reduces to the known result~\cite{majlis:92}
\begin{eqnarray}
k_\text BT_\text N=
\frac{\pi J_1}{2+\ln\left(\frac{J_1}{J_\perp}\right)}
\label{eqn:TNNAF4}.
\end{eqnarray}
This means the asymptotic $\epsilon^2\ll1$ expression for the frustrated NAF in Eq.~(\ref{eqn:TNNAF2})
can be obtained from the expression for the pure NAF by substituting $J_1\rightarrow \tilde J$ ,i.\,e., the n.n. exchange with the effective exchange of the frustrated NAF.

\subsection{CAF structure ($2J_2 >J_1$)}
\label{sec:approxCAF}

\begin{figure}
\includegraphics[width=.9\columnwidth]{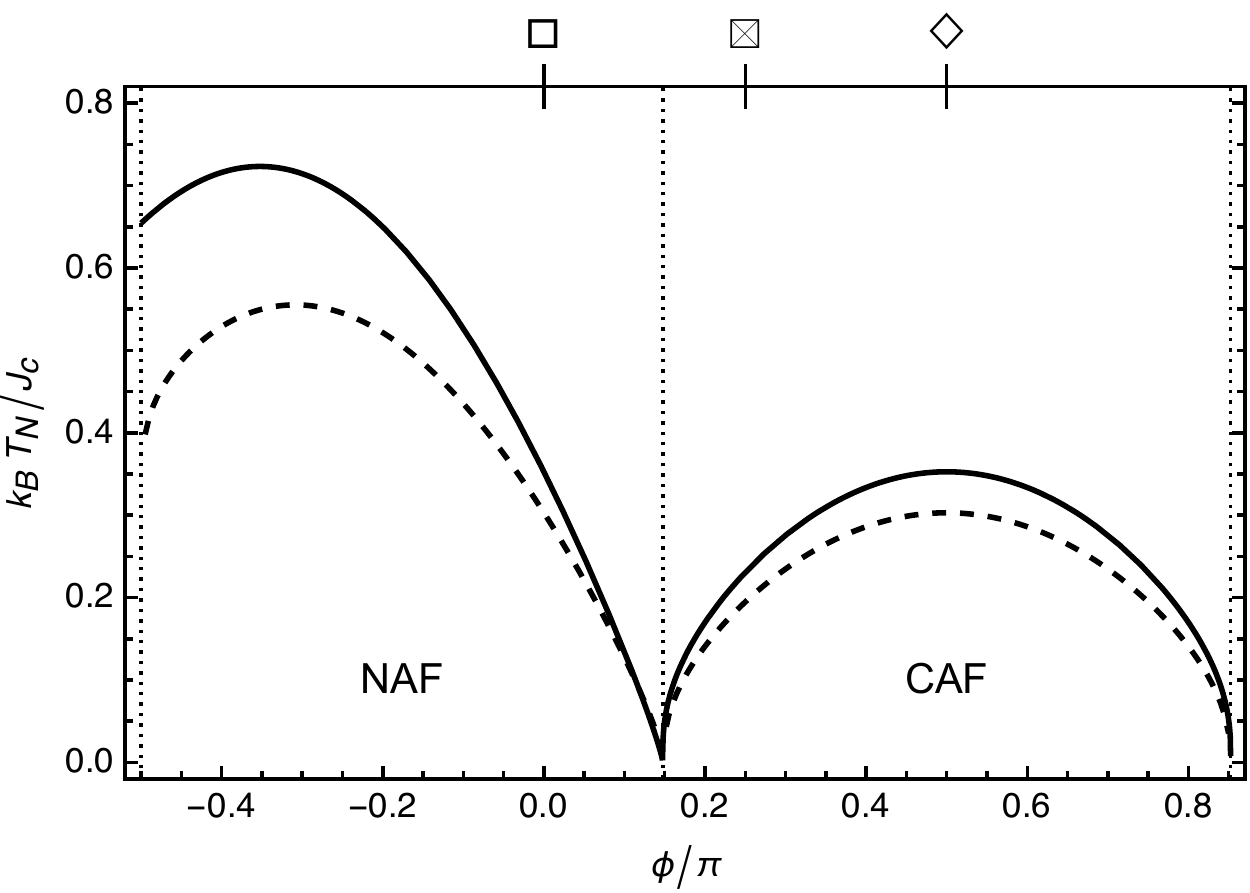}\hfill
\includegraphics[width=.9\columnwidth]{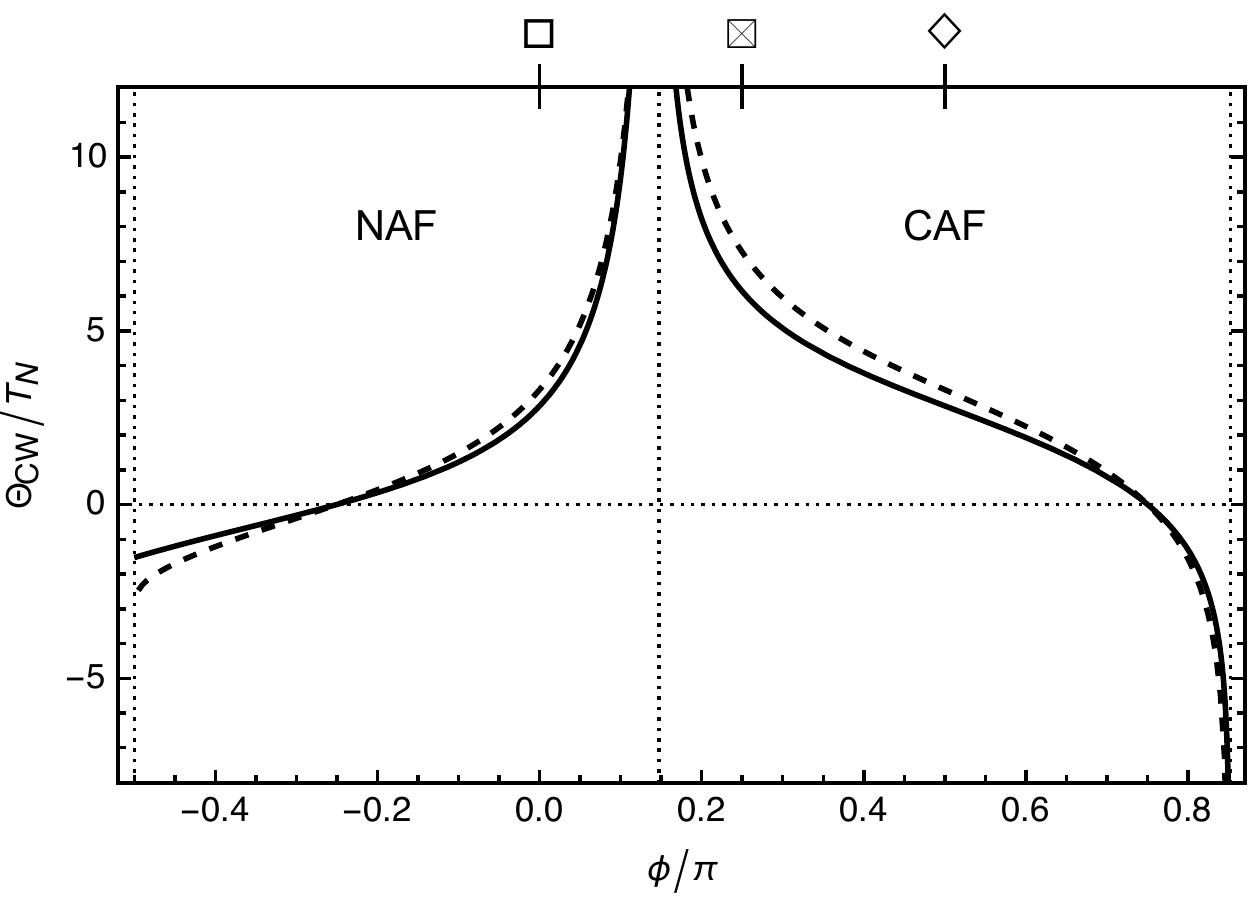}
\caption{Comparison of square lattice $T_\text N$ and $f=\Theta_\text{CW}/T_\text N$ using numerical findings from Eq.~(\ref{eqn:TN0}) (dashed lines) and analytical results (Eqs.~(\ref{eqn:TNNAF2}) and~(\ref{eqn:TNCAF1}), (full lines) results for $J_\perp/J_\text c=0.001$.}
\label{fig:An-Square}
\end{figure}
This phase breaks the fourfold in-plane symmetry, therefore the expanded dispersion $E_{\vec k}$ is not  rotationally symmetric in $\vec k_\parallel=(k_x,k_y)$. This leads to some complication because $E_{\vec k}^2$ in Eq.~(\ref{eqn:TN0})  will now depend also on $\varphi$, the azimuthal angle in $\vec k_\parallel = (k_\parallel\cos\varphi, k_\parallel\sin\varphi)$  instead of only on $k_\parallel$ as in the NAF phase (Eq.~(\ref{eqn:TNNAF1})). Therefore a final integration over $\varphi$ will remain. Furthermore the cutoff $k^\text c_\parallel$ has to be chosen such that in the CAF case $J_1=0$ which is equivalent to two decoupled interpenetrating NAF sublattices with lattice constant $\sqrt{2}a$ the same $T_\text N$ as in the previous NAF case with $J_2=0$ is obtained. Therefore $k_\parallel^\text c=\pi/\sqrt{2}$ must now be chosen. The expansion and integration in Eq.~(\ref{eqn:TN0}) then leads to 
\begin{eqnarray}
k_\text BT_\text N=(2J_2-J_1)
\frac\pi{\frac{2}{\pi}\int_0^\pi d\varphi\frac{1}{b_\varphi^2}\left[B_{\epsilon\varphi}+\ln\left(1+\frac{b_\varphi^2}{2\epsilon^2}\right)\right]}
\label{eqn:TNCAF1}
\end{eqnarray}
where we defined
\begin{eqnarray}
b_\varphi^2&=&(2J_2-J_1\cos 2\varphi)/(2J_2-J_1), \nonumber\\
B_{\epsilon\varphi}&=&\frac{b_\varphi}{\sqrt{2}\epsilon}\left(\pi-2\tan^{-1}\frac{b_\varphi}{\sqrt{2}\epsilon}\right),
\end{eqnarray}
now with $\epsilon^2=J_\perp/|\tilde J|=J_\perp/(2J_2-J_1)$. There is no simple general limiting expression for $\epsilon^2\ll 1$.
For the special case $J_1=0$ the model consists of two decoupled NAF substructures. 
Then $2\epsilon^2=J_\perp/J_2$, $b^2_\varphi =1$ and in the extreme quasi-2D case $\epsilon^2\ll 1$ we recover Eq.~(\ref{eqn:TNNAF4}) now with the replacement $J_1\rightarrow J_2$.

As stressed before these expressions contain implicitly an arbitrary momentum-cutoff (chosen as zone boundary wave number) on which the absolute value of $T_\text N$ will depend.  In the previous numerical results on the other hand the energy cutoff is given naturally by the spin wave band width. Therefore it is reasonable to compare $T_\text N$ and $f$ normalized to the unfrustrated NAF case $\phi=0$ for the two methods. For small $J_\perp/J_\text c$ gives a quite satisfactory agreement for all frustration angles $\phi$ as shown in Fig.~\ref{fig:An-Square}.
\begin{figure}
	\includegraphics[width=\columnwidth]{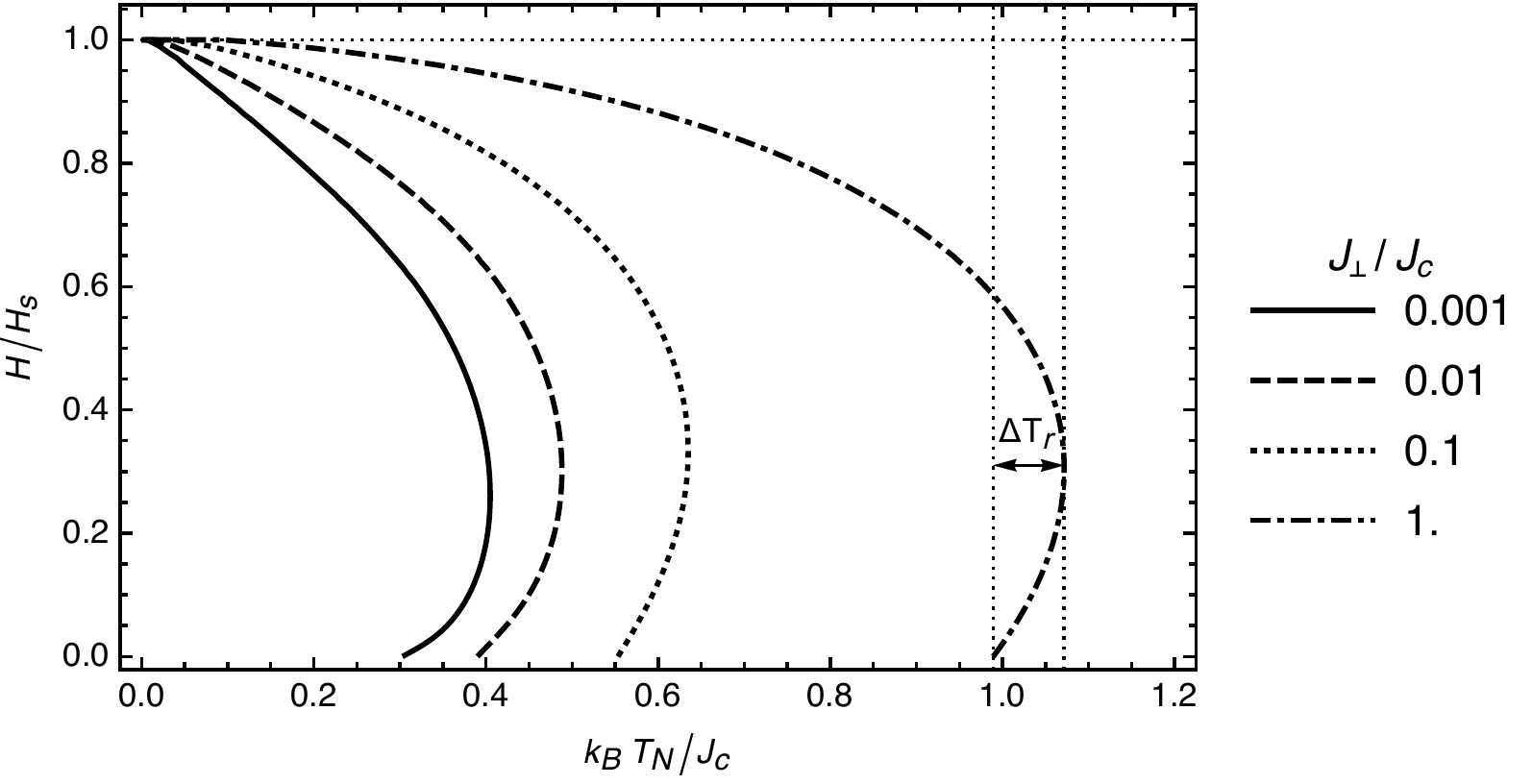}
	\caption{Field dependence of $T_\text N$ for pure quasi-2D NAF ($J_2=0$) for various interlayer coupling strengths $J_\perp/J_\text c$ as obtained from Eq.(\ref{eqn:TNH}). For $J_\perp/J_\text c=1$, the absolute reentrance difference $\Delta T_\text r$ (see text) is indicated by the small horizontal double-arrow at $H_\text{max}$.}
	\label{fig:TNH1}
\end{figure}

\section{Quasi-2D H-T phase diagram and reentrance behavior}
\label{sec:reentrance}

The ordered moment in the whole frustrated region is reduced from its classical value $m_{\vec Q}=S$ by a considerable amount (e.\,g. $0.606S$ in the pure N\'eel case of the square lattice)~\cite{schmidt:11,schmidt:17}. It was shown before, using LSW and ED approach~\cite{siahatgar:11}  that the application of a magnetic field strongly reduces the quantum fluctuations. Therefore initially, for a small applied field $B=\mu_0H$ the ordered moment increases and on approaching the saturation field $H_\text s$ decreases again due to the classical geometric canting effect, leading to a nonmonotonic behavior of $m_{\vec Q}(H)$ which was observed~\cite{tsyrulin:10} and explained~\cite{siahatgar:11} for the quasi-2D $S=1/2$ quantum magnet Cu(pz)$_2$(ClO$_4$)$_2$~\cite{tsyrulin:09}.

The nonmonotonic behavior of  $m_{\vec Q}(H)$ is most pronounced for strong frustration, i.\,e. when the initial value  $m_{\vec Q}(0)$ is strongly suppressed. A complementary effect is seen in the field dependence of the ordering temperature $T_{\text N}(H)$~\cite{tsyrulin:10}. For small fields $H\ll H_\text s$ it was shown to increase~\cite{siahatgar:11}, again due to the reduction of quantum fluctuations by the applied field. For larger fields approaching the saturation value $H_\text s$, $T_{\text N}(H)$ eventually has to vanish. Thus, due to the initial increase of $T_{\text N}(H)$ a reentrance behavior of the magnetic order as function of the applied field at constant temperature $T>T^0_{\text N}$ has to be expected.

\begin{figure}
\includegraphics[width=\columnwidth]{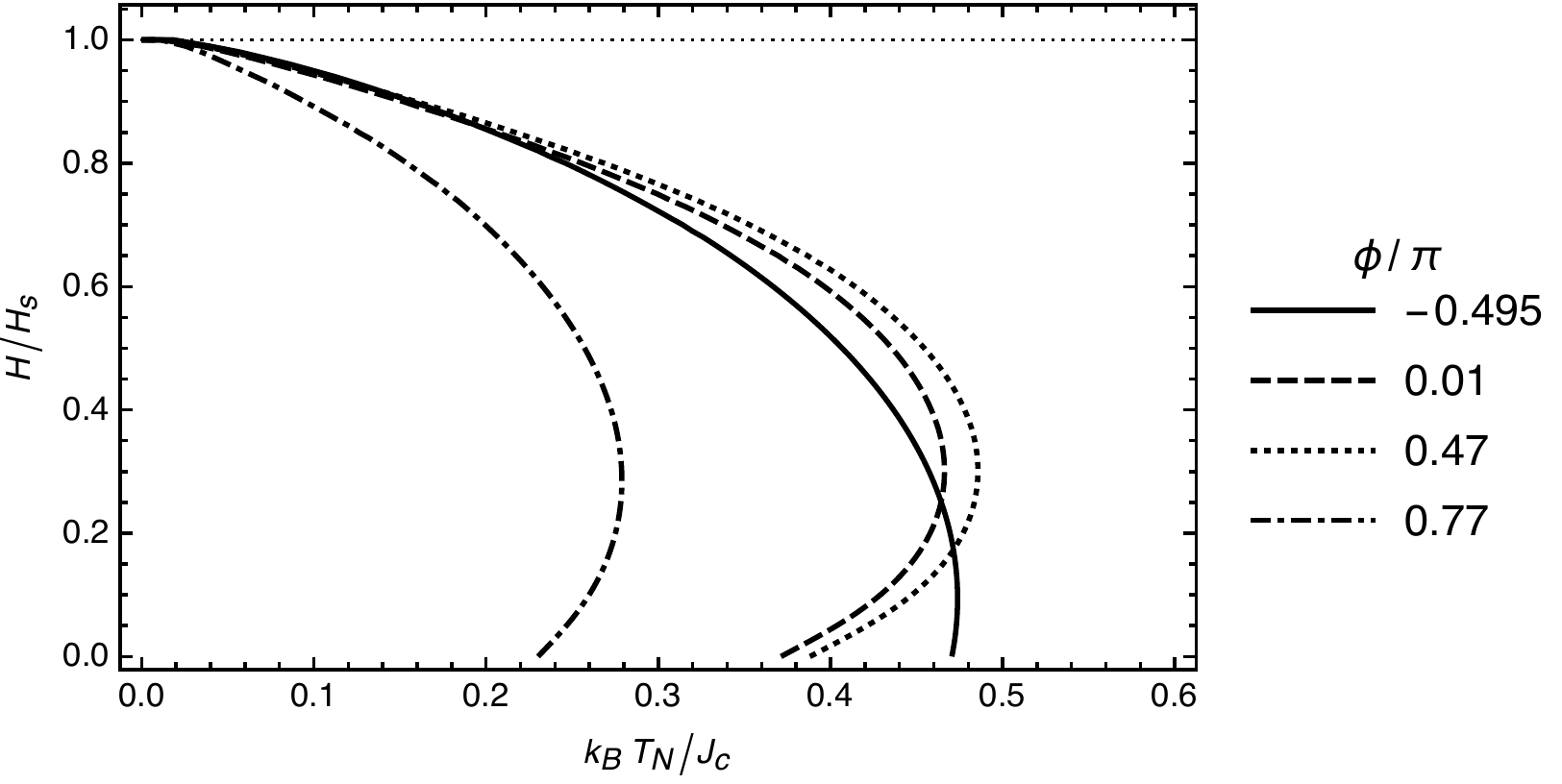}
\caption{Field dependence of $T_\text N$ for interlayer coupling strength $J_\perp/J_\text c=0.01$ and various
$\phi$-values corresponding to NAF/FM boundary (full line) and various compound values listed in Table~\ref{tbl:exchange}.
The former shows no reentrance due to absence of quantum fluctuations for $\phi/\pi\simeq -0.5$.}
\label{fig:TNH2}
\end{figure}
\begin{figure}
\includegraphics[width=.9\columnwidth]{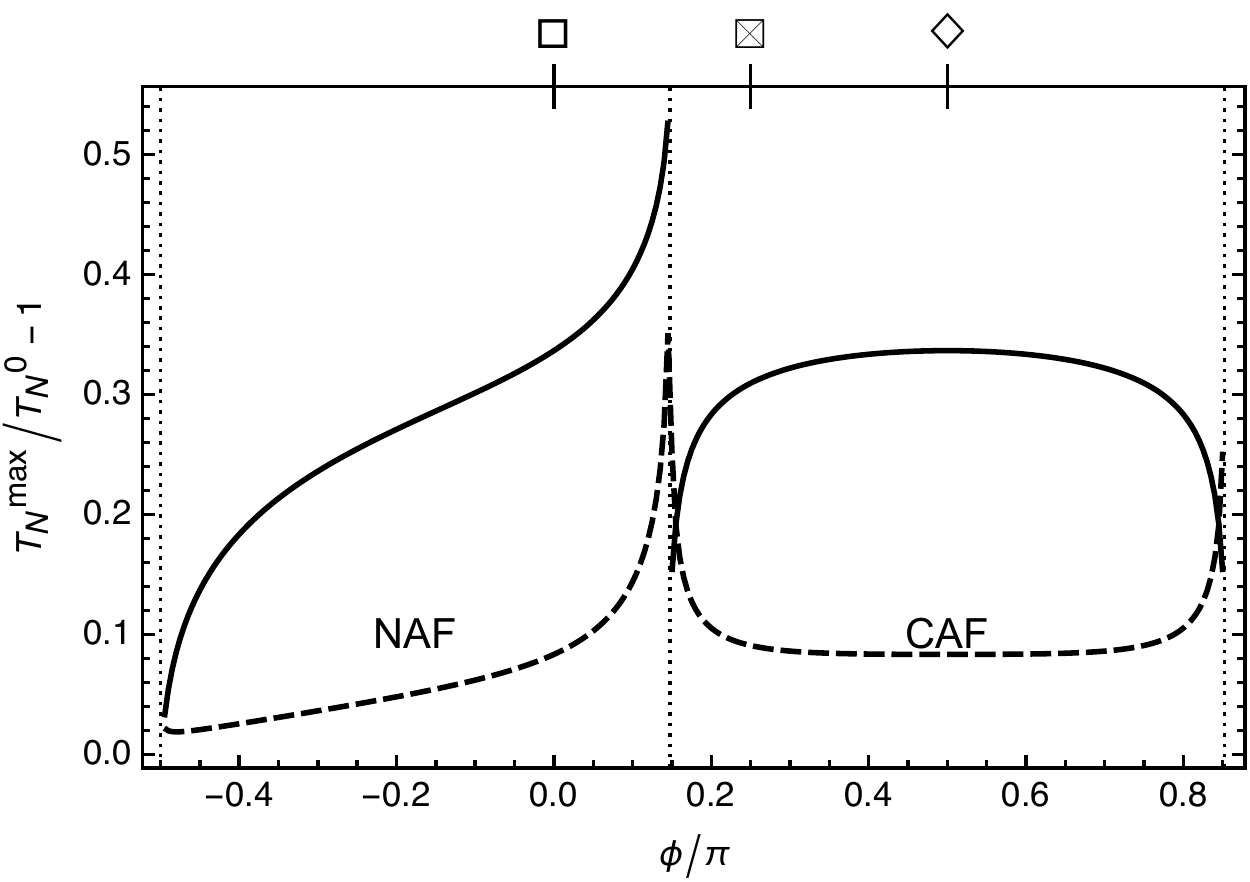}
\caption{Reentrance measure given by the difference between the field-dependent maximal Néel temperature and its zero-field value normalized 
to the latter, $\delta T_\text r=T_\text N^\text{max}/T_\text N^0-1$ versus frustration angle $\phi$ for interlayer coupling strengths 
$J_\perp/J_\text c=0.001$ (solid line) and $J_\perp/J_\text c=1$ (dashed line).}
\label{fig:tnmax}
\end{figure}
The full phase diagram for all fields and frustration ratios of a quasi-2D magnet  is investigated  in the present section. It is obtained from the iterative solution of Eq.~(\ref{eqn:TNH}) which provides us with the phase boundary $T_{\text N}(H)$ for the quasi-2D magnet for all parameter sets $(\phi,J_\perp/J_\text c)$. First we consider the pure N\'eel case $(\phi = 0)$ shown in Fig.~\ref{fig:TNH1}. The reentrance behavior caused by the field dependence of quantum fluctuations is clearly seen. The absolute reentrance difference of maximum and zero-field ordering temperature $\Delta T_\text r=T_\text N^\text{max}(H_\text{max})-T^0_N$ is rather independent of the 2D character. However, the relative difference $\delta T_\text r=\Delta T_\text r/T_\text N^0= T_\text N^\text{max}/T_\text N^0-1$ which is a measure for the prominence of reentrance in the transition line increases with decreasing $J_\perp/J_\text c$.

In the complementary Fig.~\ref{fig:TNH2} we show the frustration $(\phi)$ dependence of the transition line for an intermediate 2D character with $J_\perp/J_\text c=0.01$. Close to the NAF/FM boundary $\phi/\pi=-0.49$ where quantum fluctuations are strongly reduced the reentrance behavior characterized by $\delta T_\text r$ vanishes. It increases rapidly in the whole unfrustrated $\phi<0$ regime which proves that the reentrance is primarily associated with the field-dependent suppression of quantum fluctuations and not so much with the effect of frustration. It does, however achieve a maximum on approaching the strongly frustrated regime $\phi/\pi=0.15$. This is most clearly seen when we plot the reentrance measure  $\delta T_\text r$ as function of $\phi$ (Fig.\ref{fig:tnmax}) for extreme 2D case (full line) and isotropic 3D case (dashed line). In the NAF case indeed  $\delta T_\text r$ increases monotonically from NAF/FM $(\phi/\pi=-0.5)$ to NAF/CAF  $(\phi/\pi=0.15)$ boundaries. In the main part of the CAF phase it stays almost at constant value equal to that of the unfrustrated case. Generally  $\delta T_\text r$ is much larger in the extreme quasi-2D magnet (full line) for both phases. Interestingly in this case the reentrance measure decreases when approaching the strongly frustrated phase boundaries from the CAF side.

\section{Application to quasi-2D oxovanadate compounds}
\label{sec:oxo}

\begin{table*} 
	\begin{tabular}[c]{l|cccccccccc|cc}
	Compound & $\phi/\pi$ & $J_\text c/(k_{\text B}\text K)$ &
	$J_1/(k_{\text B}\text K)$  & $J_2/(k_{\text B}\text K)$ &
	$|\tilde J|/(k_{\text B}\text K)$ & 
	$\Theta_\text{CW}/{\text K}$ & $T_{\text N}/\text K$  &
	$ k_{\text B}T_{\text N}/J_\text c$ &  $k_{\text B}T_{\text N}/|\tilde J|$ & 
	$f$ & Ref.&Symb.  \\
	\hline
	Zn$_{2}$VO(PO$_4$)$_2$ & 0.008 & 7.9 &
	7.91 & 0.2 &7.5 & 8.11 &  3.7 & 0.46 & 0.49 & 2.19 & 
	\cite{kini:06}&$\bullet$\\
	Li$_2$VOGeO$_4$ & 0.44 & 4.2 &
	0.82 & 4.1& 7.38 & 4.92 & 2.1 & 0.50 & 0.28 & 2.34 &
	\cite{kaul:04,kaul:05} &$\triangle$\\
    Li$_2$VOSiO$_4$ & 0.47 & 6.3 &
    0.56 & 6.3 & 12.04 & 6.86 & 2.7 & 0.43 & 0.22 & 2.54 &
    \cite{kaul:04,kaul:05}&$\triangle$\\
    Pb$_2$VO(PO$_4$)$_2$ & 0.60 & 6.8 &
    -2 & 6.5 & 16.25 & 4.5 & 3.5 & 0.51 & 0.21 & 1.28 &
    \cite{skoulatos:07}&$\blacklozenge$\\
    Pb$_2$VO(PO$_4$)$_2$ & 0.63 & 8.4 &
    -3.2 & 7.7 & 18.8 & 4.5 & 3.7 & 0.44 & 0.20 & 1.22 &
    \cite{skoulatos:09}&$\blacklozenge$\\
    PbZnVO(PO$_4$)$_2$ & 0.65 & 11.27 &
    -5.2 & 10.0 & 25.2 & 4.8 & 3.9 & 0.35 & 0.15 & 1.23 &
    \cite{tsirlin:10}&$\square$\\
    Na$_{1.5}$VO(PO$_4$)$_2$F$_{0.5}$ & 0.65 &7.1&
    -3.2 & 6.3 & 15.8 & 3.1 & 2.6 & 0.36 & 0.16 & 1.19 &
    \cite{tsirlin:09}&$\square$\\
    BaZnVO(PO$_4$)$_2$ & 0.66 & 10.5 &
    -4.99 & 9.26 & 23.5 & 4.27 & 3.8 & 0.36 & 0.16 & 1.12 &
    \cite{kaul:05}&$\square$\\
	Pb$_2$VO(PO$_4$)$_2$ & 0.66 & 10.7 &
	-5.1 & 9.4 & 23.9 & 4.3 & 3.7 & 0.35 & 0.15 & 1.16 &
	\cite{tsirlin:09}&$\blacklozenge$\\
	Pb$_2$VO(PO$_4$)$_2$ & 0.67 & 11.5 &
	-6 & 9.8 & 25.6 & 3. 8 & 3.7 & 0.32 & 0.14 & 1.02 &
	\protect\cite{kaul:04}&$\blacklozenge$\\
	SrZnVO(PO$_4$)$_2$ & 0.73 & 12.2 &
	-8.3 & 8.9 & 26.1 & 0.6 & 2.7 & 0.22 & 0.10 & 0.22 &
	\cite{tsirlin:09}&$\square$\\
    BaCdVO(PO$_4$)$_2$ & 0.77 & 4.8 &
    -3.6 & 3.2 & 10.0 & -0.4 & 1.0 & 0.21 & 0.10 & -0.4 &
    \cite{nath:08,tsirlin:09}&$\square$\\
	\end{tabular}
    \caption[Exchange interactions constants for vanadium oxide compounds]
    {Exchange interactions constants for various vanadium oxide compounds, ordered with increasing $\phi$, i.\,e. approaching the CAF/FM boundary. Results are obtained mostly from susceptibility $\chi(T)$ and magnetization $m_0(h)$ analysis, except for fourth and fifth row which are deduced from neutron diffraction. Here $\Theta_\text{CW} =(J_1+J_2)/k_{\text B}$ is the 2D Curie-Weiss temperature, $T_{\text N}$ the N\'eel temperature and $f$ the empirical frustration ratio. The symbols in the last column are used in Fig.~\ref{fig:tn_Pb2} to label the $k_\text BT_\text N/J_\text c$ data points shown there. The example of  Pb$_2$VO(PO$_4$)$_2$ shows that a certain variation in exchange parameters as determined by different methods and in different references occurs.}
     \label{tbl:exchange}
\end{table*}

\begin{figure}
\includegraphics[width=.9\columnwidth]{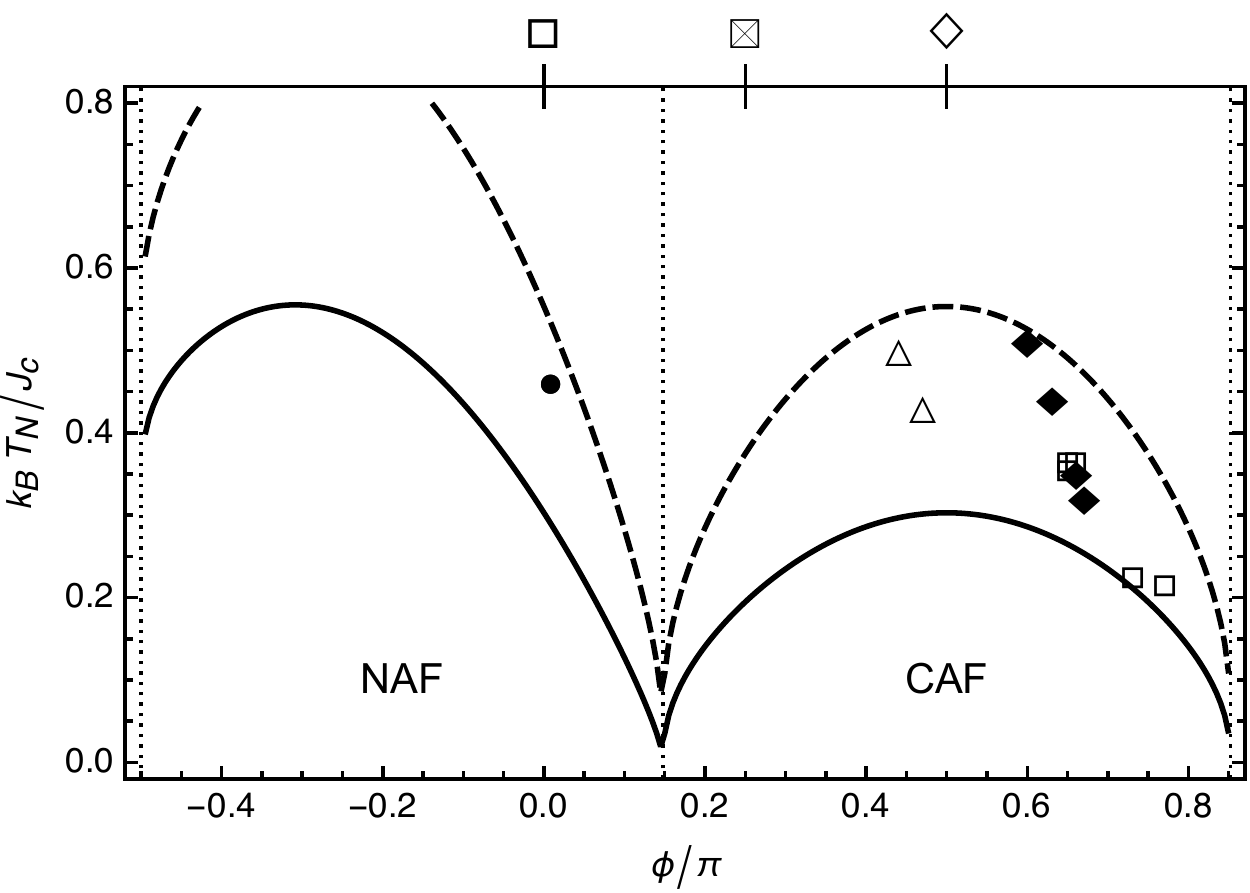}
\caption{Néel temperature $T_\text N$ versus frustration angle $\phi$. Dots: experimental values according to
 Table~\ref{tbl:exchange}. Lines from evaluation of Eq.~(\ref{eqn:TN0}) for $J_\perp/J_\text c=0.001$ (full) 
 and~$0.1$ (dashed).}
\label{fig:tn_Pb2}
\end{figure}

The discovery of two classes of layered vanadium oxides Li$_2$VO$X$O$_4$ ($X=\text{Si},\text{Ge}$)~\cite{millet:98,melzi:00,melzi:01,carretta:04} and $AA'$VO(PO$_4$)$_2$ ($A,A'=\text{Pb},\text{Zn},\text{Sr},\text{Ba}$)~\cite{kaul:04,kini:06,nath:08,nath:09} provided a variable platform of 2D frustrated quantum magnets with different chemical composition. Nevertheless their magnetism is described universally by the $J_1$-$J_2$ model with $J_2/J_1$ or $\phi$ depending on the specific compound. Each of them features V$^{4+}$ ions with $S=1/2$ surrounded by oxygen polyhedra, forming layers of $J_1$-$J_2$ square lattices with weak interlayer coupling~\cite{melzi:01,kini:06}.

These compounds  were experimentally investigated e.\,g. in Refs.~\onlinecite{kaul:04,skoulatos:07,skoulatos:09} using susceptibility and specific heat measurements as well as  neutron diffraction as tools. The typical observed signatures in these experiments point to quasi-2D magnetism. However, the actual size of $J_\perp/J_\text c$, i.\,e. the inter/intra-layer exchange ratio has not been estimated  because an applicable theory for $T_\text N(\phi,J_\perp)$ for all $\phi$ was lacking. Using the theoretical results of the previous sections we can now give an assessment of the size of $J_\perp$ within the series. As input we use the $J_1$ and $J_2$ values obtained previously from 2D finite-temperature Lanczos method (FTLM) applied to the experimental susceptibilities or neutron diffraction results~\cite{schmidt:17}  and the experimental values of $T_{\text N}$ listed among other items in Table~\ref{tbl:exchange}.

Such a thermodynamic analysis can give only some estimated range of $J_\perp/J_\text c$ because on the one hand the experimental value of $k_\text BT_\text N/J_\text c$ may be rather uncertain. For example for Pb$_2$VO(PO$_4$)$_2$, it varies between $0.32$ and $0.51$ ($\blacklozenge$~in Table \ref{tbl:exchange}).  On the other hand the Néel temperature depends only logarithmically on $J_\perp$  (Eqs.~(\ref{eqn:TNNAF2},\ref{eqn:TNCAF1})) and therefore a wide range of values for $J_\perp/J_\text c$ is possible. We plot the experimental values of $k_\text BT_\text N/J_\text c$  from  Table \ref{tbl:exchange} together with two theoretical curves in Fig.~\ref{fig:tn_Pb2}. The experimental values all lie in a corridor limited by the theoretical results for   $J_\perp/J_\text c=0.1$ (full line) and  $J_\perp/J_\text c=0.001$ (dashed line). We conclude that the oxovanadate series are indeed quasi-2D magnets but with non-negligible interlayer coupling.

The field dependence of the N\'eel temperature caused by the suppression of fluctuations is discussed in Sec.~\ref{sec:reentrance}. In the example of Cu(pz)$_2$(ClO$_4$)$_2$~\cite{tsyrulin:10,siahatgar:11} it was observed and calculated. However due to the rather high exchange energy scale $J_\text c/k_\text B=18.6\,\rm K$ the field for the maximum Néel temperature $T_{\text N}^\text{max}(H)$ is not reached such that the phase diagram with reentrance character, although certainly present, has not been fully determined.

A more favorable case is Pb$_2$VO(PO$_4$)$_2$~\cite{kaul:05}. The smaller exchange energy scale (Table \ref{tbl:exchange}) makes it possible to reach $T_{\text N}^\text{max}(H)\approx3.9\,\rm K$ at $\mu_0H\approx8\,\rm T$ in susceptibility and specific heat measurements~\cite{kaul:04}. At the largest accessible field $\mu_0H=14\,\rm T$ the  $T_\text N(H)$ curve has started to turn back. The experimental values  together with the optimal theoretical curve for $J_\perp/J_\text c=0.02$ 
and $\phi/\pi=0.63$ appropriate for Pb$_2$VO(PO$_4$)$_2$ are plotted in Fig.~\ref{fig:reent_Pb2}.

The fitting of the whole $T_{\text N}(H)$ curve leads to a more reliable value for $J_\perp$ than just comparing $T^0_{\text N}$ as in Fig.~\ref{fig:tn_Pb2}. The observed experimental reentrance behavior is somewhat less pronounced than expected for these parameters. One reason certainly is that Pb$_2$VO(PO$_4$)$_2$ has a small Ising anisotropy that suppresses part of the fluctuations and therefore $T_\text N^0$ is less reduced by quantum fluctuations than in the pure isotropic Heisenberg case. The presence of this Ising term may be concluded from a spin-flop transition below $\mu_0H_\text{sf}\approx0.9\,\rm T$ not shown in the data of Fig.~\ref{fig:reent_Pb2}. 

In any case one may expect that reentrant behavior is a ubiquitous phenomenon for quasi-2D quantum magnets due to the universal mechanism of suppressing quantum fluctuations by application of the field. This mechanism is also obvious from the temperature dependence of the specific heat, for example in  Pb$_2$VO(PO$_4$)$_2$ which shows an increasing sharpening of the transition peak for increasing field~\cite{kaul:05} due to the suppression of spin fluctuations, a very typical behavior for reentrant phase transitions. The present theory may also be applied to study this effect.

\section{Discussion and Summary}
\label{sect:summary}

We have investigated the frustration, interlayer coupling and field dependence of the N\'eel temperature in quasi-2D quantum magnets.
Based on the simple Tyablikov self-consistency modification of LSW theory we have derived equations that should qualitatively describe
the systematic variation of $T_\text N$ with frustration parameter $\phi=\tan^{-1}(J_2/J_1)$, interlayer coupling $J_\perp$ and the applied field $H$ up to the saturation value $H_\text s$. Furthermore we investigated to what extent the experimentally used empirical frustration ratio $f=\Theta_\text{CW}/T_\text N$ 
(which may be positive or negative) is a relevant measure of frustration. In the mean field approximation $|f|$ is of order one or less.

We find that indeed the N\'eel temperature $T_\text N(\phi)$ is strongly suppressed for regions with large 
frustration and accordingly $|f|$ may be greatly enhanced in narrow intervals around these special points with $|2J_2/J_1|=1$ or $\phi/\pi\approx0.15,0.85$ where the ground state magnetic moment breaks down and spin liquid or spin nematic states~\cite{schmidt:17} are established. In these regions $|f|\gg 1$ can function as a useful measure of frustration, in particular in the square lattice (Fig.~\ref{fig:square_TNf}). The logarithmic dependence of $T_\text N$ on interlayer coupling $J_\perp$ for the unfrustrated case $J_2=0$ is confirmed to hold in both NAF and CAF sectors for all frustration degrees. This is also proven by explicit analytical approximations for $T_\text N(\phi)$. 
\begin{figure}
\includegraphics[width=\columnwidth]{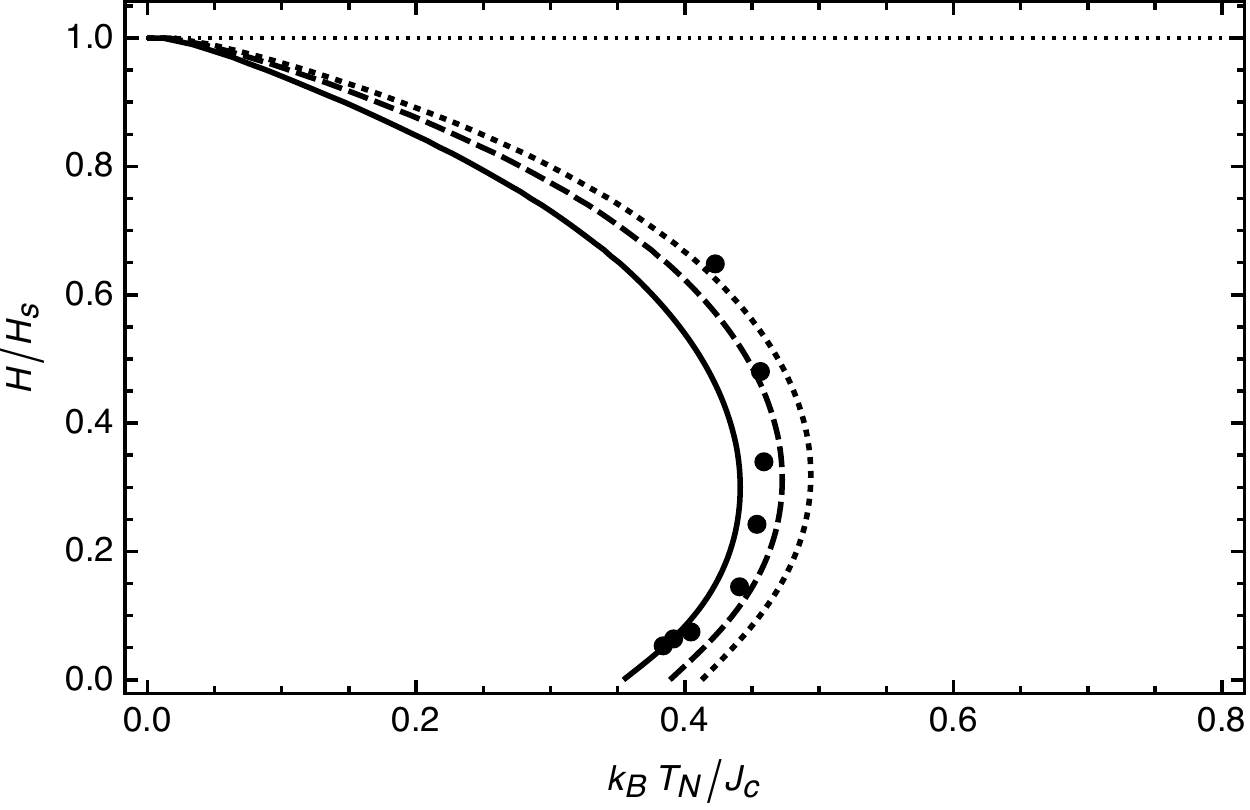}
\caption{Reentrance behavior of $T_\text N(H)$ in  Pb$_2$VO(PO$_4$)$_2$~\cite{kaul:05} (black dots). 
Good agreement is obtained for $J_\perp/J_\text c= 0.02$ (dashed line) using $\phi/\pi=0.63$ and $J_\text c=8.4\,\text K/k_\text B$ from Table~\ref{tbl:exchange} and a saturation field $\mu_0H=20.9\,\text T$~\cite{schmidt:17}.
Full and dotted curve correspond to values $J_\perp/J_\text c=0.01,0.03$ respectively.}
\label{fig:reent_Pb2}
\end{figure}

However our analysis shows that  the criterion $|f|\gg1$ as an indicator for large frustration may also be misleading. Especially in the anisotropic triangular lattice (Fig.~\ref{fig:triangular_TNf}) the criterion is also fulfilled for the {\em unfrustrated\/} quasi-1D cases with $\phi/\pi=\pm 0.5$, being far away from the special regions of strong frustration with $\phi/\pi\approx0.15$ and~$0.85$ (see Fig.~\ref{fig:frustration}). Instead critical 1D fluctuations lead to the enhancement of $f$ here. Particular examples are the well known anisotropic triangular magnets CsCuCl$_4$ ($f=5.6$) and  CsCuBr$_4$ ($f=9.2$). They have strongly enhanced $f$ values but anisotropy parameters $\phi/\pi=0.41$ and~$0.38$, respectively~\cite{coldea:03,starykh:10,schmidt:15}. Although they are still quasi-2D magnets with finite $T_\text N$ of $0.62\,\text K$ and $1.42\,\text K$, respectively, they are already placed close the quasi-1D region of the phase diagram with interchain couplings $J_1/J_2 \approx0.29$ and~$0.40$ and rather reduced frustration degree (Fig.~\ref{fig:frustration}). That quasi-1D fluctuations are the reason for large $f$-values in these compounds is also directly evident from the typical quasi-1D spinon excitation continuum observed in inelastic neutron scattering experiments~\cite{coldea:03}.

Furthermore the large family of square lattice oxovanadate quasi-2D magnets with one exception have frustration angles in the interval $0.44<\phi/\pi<0.77$. This begins close to the unfrustrated CAF and ends before the strongly frustrated CAF region slightly below $\phi/\pi\approx0.85$. Therefore the values of $|f|$ in this family of compounds are rather close to one (Table~\ref{tbl:exchange}) without dramatic variation.

In essence then, be it the square or anisotropic triangular lattice, in order to use the size of $f=\Theta_\text{CW}/T_\text N$  as a frustration criterion for a compound investigated one should have additional information beforehand about its location in the phase diagram. This can for example be the  determination of $\phi$ by a FTLM fit to the temperature dependence of the magnetic susceptibility.

We have analyzed the field dependence of the N\'eel temperature and found it is determined by the universal effect of reduction of moment fluctuations by the applied field. It is known that this mechanism leads to a non-monotonic field dependence of the ordered moment~\cite{tsyrulin:10,siahatgar:11}. The present analysis has shown that it also leads to a non-monotonic $T_\text N(H)$ behavior, i.\,e. a reentrance character of the $H$-$T$ phase diagram in quantum magnets. The quantity $\delta T_\text r=\Delta T_\text r/T_\text N^0=T_\text N^\text{max}/T_\text N^0-1$ characterizing the reentrance shows a pronounced dependence on frustration angle in the strongly frustrated regimes around the classical phase boundaries NAF/CAF and CAF/FM. Outside these regions it increases strongly with decreasing $J_\perp$ on approaching the extreme quasi-2D limit.

Such reentrance phase diagrams as in Figs.~\ref{fig:TNH2} and~\ref{fig:reent_Pb2} should therefore be ubiquitous among quasi-2D magnets but  may not always easily be observable in the experimentally available range of magnetic fields. The oxovanadate Pb$_2$VO(PO$_4$)$_2$ is an exception where reentrance has been found due to a modest estimated saturation field $\mu_0H_\text s\approx20.9\,\rm T$, a consequence of the relatively small exchange constants $J_1$ and $J_2$ of the material, see Table~\ref{tbl:exchange}. $T_\text N(H)$ follows qualitatively the expected behavior, however the details may be subject to exchange anisotropies not taken into account here and further interactions which influence zero-field spin fluctuations and thus modify the zero-field value $T_\text N^0$.  The inclusion of such effects in the present framework via modified spin wave excitations seems rather straightforward.

\section*{Acknowledgements} We would like to thank Ch. Geibel for discussion and the permission to use unpublished data.

\appendix
\section{Global and local spin coordinates in a magnetic field}
\label{sect:app1}
The spin wave approximation is performed in a coordinate system where the 
local $z$ direction at a given site $i$ coincides with the moment direction at that site. The connection to the global spin coordinates used in Eq.~(\ref{eqn:hex12}) is given by
\begin{widetext}
\begin{equation}
    \left(
    \begin{array}{c}
  S_{i}^{x}\\
  S_{i}^{y}\\
  S_{i}^{z}
    \end{array}
    \right) =
    \left(
    \begin{array}{ccc}
  \cos(\vec Q\vec R_{i}) & -\sin(\vec Q\vec R_{i}) & 0\\
  \sin(\vec Q\vec R_{i}) & \cos(\vec Q\vec R_{i}) & 0\\
  0 & 0 & 1
    \end{array}
    \right)
    \left(
    \begin{array}{ccc}
  \cos\Theta & 0 & \sin\Theta\\
  0 & 1 & 0\\
  -\sin\Theta & 0 & \cos\Theta
    \end{array}
    \right)
    \left(
    \begin{array}{c}
  S_{i}^{x'}\\
  S_{i}^{y'}\\
  S_{i}^{z'}
    \end{array}
    \right).
\end{equation}
\end{widetext}
For clarity we denote the local spin coordinates with primes in this expression. While the first matrix represents the in-plane $xy$ rotation due to spontaneous order characterized by $\vec Q$ the second one describes the $xz$-plane canting of spins with angle $\Theta$ caused by the magnetic field with a classical value $\cos\Theta_\text{cl}=H/H_\text s$.

\section{Expansion integrals for the ordered moment}
\label{sect:app2}

In zero field for $T\leq T_\text N$ the approximate solution for the total moment $\langle S\rangle$ in Eq.~(\ref{eqn:neel:sbarmf3}) is determined by expansion coefficients that are expressed in terms of frustration ($\phi$) dependent integrals $I_0\ldots I_2$. They are given by
\begin{eqnarray}
	\label{eqn:neel:i}
	I_0
	&:=&
	\frac13\int_\text{BZ}\frac{{\rm d}^3k}{V_\text{BZ}}
	\left(A_{\vec k}-B_{\vec k}\cos\Theta_\text{cl}\right),
	\\
	I_1
	&:=&
	\frac1{180}\int_\text{BZ}\frac{{\rm d}^3k}{V_\text{BZ}}
	\left(A_{\vec k}-B_{\vec k}\cos\Theta_\text{cl}\right)
	E^2_{\vec k},
	\\
	I_2
	&:=&
	\frac1{7560}\int_\text{BZ}\frac{{\rm d}^3k}{V_\text{BZ}}
	\left(A_{\vec k}-B_{\vec k}\cos\Theta_\text{cl}\right)
	E^4_{\vec k}.
\end{eqnarray}
Using these integrals in Eq.~(\ref{eqn:neel:sbarmf3}) leads to the approximate $\langle S\rangle(T)$ curves in Fig.~\ref{fig:msapprox}.
\vspace*{.5\baselineskip}

\section{Néel temperature at the CAF borders}
\label{sec:tncafborders}

From Eq.~(\ref{eqn:TN0}), we obtain
\begin{eqnarray}
	\beta_\text N
	&=&
	4\int_\text{BZ}\frac{{\rm d}^3k}{V_\text{BZ}}
	\frac{A_{\vec k}}{A^2_{\vec k}-B^2_{\vec k}}
	\nonumber\\
	&=&
	2\int_\text{BZ}\frac{{\rm d}^3k}{V_\text{BZ}}
	\left(
	\frac1{A_{\vec k}-B_{\vec k}}+\frac1{A_{\vec k}+B_{\vec k}}
	\right)
	\nonumber\\
	&=&
	4\int_\text{BZ}\frac{{\rm d}^3k}{V_\text{BZ}}
	\frac1{A_{\vec k}+B_{\vec k}}
	\label{eqn:betancaf}
\end{eqnarray}
where the last equality holds for commensurate ordering vectors $\vec Q$ with components $(Q_x,Q_y,Q_z)=\pi(n_x,n_y,n_z)$ and $n_x,n_y,n_z\in \mathbb Z$. At the CAF borders, we have either $J_2=J_1/2$ (border to NAF) or $J_2=-J_1/2>0$ (border to FM) such that we obtain from Eq.~(\ref{eqn:swcoeff})
\begin{equation}
	A_{\vec k}+B_{\vec k}
	=
	2\left[
	J_1(1+\cos k_x)(1+\cos k_y)+J_\perp(1+\cos k_z)
	\right]
\end{equation}
at the NAF/CAF border and a similar expression at the FM/CAF border. The denominator in Eq.~(\ref{eqn:betancaf}) therefore has lines of zeroes at $(k_x,k_y,k_z)=(\pi,k_y,\pi)$ or $(k_x,k_y,k_z)=(k_x,\pi,\pi)$, implying that the integral~(\ref{eqn:betancaf}) diverges and $T_\text N\to0$ eventually.

\bibliography{references}

\end{document}